\documentclass[12pt]{article}
\pdfoutput=1
\usepackage{graphicx}
\usepackage{epstopdf}
\usepackage{amsmath}
\usepackage{amsfonts}
\usepackage{amssymb}
\usepackage{color}
\usepackage{mathrsfs}

\newcommand{\be}{\begin{equation}}
\newcommand{\ee}{\end{equation}}
\newcommand{\bea}{\begin{eqnarray}}
\newcommand{\eea}{\end{eqnarray}}
\newcommand{\barr}{\begin{array}}
\newcommand{\earr}{\end{array}}

\usepackage[colorlinks,bookmarks]{hyperref}
\definecolor{linkblue}{rgb}{0,0,0.8}
\definecolor{linkgreen}{rgb}{0,0.5,0}

\hypersetup{pdfpagemode=UseNone, pdfstartview=FitH, linkcolor=linkblue, %
            citecolor=linkgreen, urlcolor=linkblue}

\def\beq{\begin{equation}}
\def\eeq{\end{equation}}
\def\be{\begin{equation}}
\def\ee{\end{equation}}
\def\bea{\begin{eqnarray}}
\def\eea{\end{eqnarray}}
\def\d{{\partial}}

\def\nn{\nonumber}

\def\kreach{{k_{\rm reach}}}
\def\knl{{k_{\rm NL}}}

\def\cstoch{c_{\rm stoch}}

\def\co{c_{s  (1)}^2}
\def\ct{c_{s  (2)}^2}

\newcommand{\kren}{k_\text{ren}}

\newcommand{\invMpc}{\,h\, {\rm Mpc}^{-1}\,}
\newcommand{\hinvMpc}{\,h\, {\rm Mpc}^{-1}\,}

\newcommand{\lp}{\left(}
\newcommand{\rp}{\right)}
\newcommand{\lb}{\left[}
\newcommand{\rb}{\right]}

\newcommand{\vq}{\vec{q}}

\newcommand{\kmax}{k_{\rm max}}
\newcommand{\kfit}{k_{\rm fit}}
\newcommand{\gammai}{(\d\tau)_{\rho_l}}

\def\poneloop{P_\text{1-loop}}
\def\ptwoloop{P_\text{2-loop}}

\def\ptreecs{P_\text{tree}^{(c_{\rm s})}}

\def\poneloopcs{P_\text{1-loop}^{(c_{\rm s})}}

 \definecolor{purple}{rgb}{0.78,0.18,0.77}
\def\cpur{\bf \color{purple}}

\begin{document}


\setcounter{page}{1} \baselineskip=15.5pt \thispagestyle{empty}

\begin{flushright}
\end{flushright}

\begin{center}

{\Large \bf Precision Comparison of \\[0.5cm] the Power Spectrum in the EFTofLSS with Simulations}
\\[0.7cm]
{\large Simon Foreman${}^{1,2}$, Hideki Perrier${}^{3}$ and Leonardo Senatore${}^{1,2}$ }
\\[0.7cm]
{\normalsize { \sl $^{1}$ Stanford Institute for Theoretical Physics,\\ Stanford University, Stanford, CA 94306}}\\
\vspace{.3cm}

{\normalsize { \sl $^{2}$ Kavli Institute for Particle Astrophysics and Cosmology, \\
SLAC and Stanford University, Menlo Park, CA 94025}}\\
\vspace{.3cm}

{\normalsize { \sl $^{3}$ University of Geneva, Department of Theoretical Physics \\
 and Center for Astroparticle Physics (CAP), \\
 24 quai E. Ansermet, CH-1211 Geneva 4, Switzerland}}\\
\vspace{.3cm}

\end{center}

\vspace{.8cm}

\hrule \vspace{0.3cm}
{\small  \noindent \textbf{Abstract} \\[0.3cm]
\noindent  
We study the prediction of the dark matter power spectrum at two-loop order in the Effective Field Theory of Large Scale Structures (EFTofLSS) using high precision numerical simulations. In our universe, short distance non-linear fluctuations, not under perturbative control, affect long distance fluctuations through an effective stress tensor that needs to be parametrized in terms of counterterms that are functions of the long distance fluctuating fields.  We find that at two-loop order it is necessary to include three counterterms: a linear term in the overdensity,~$\delta$, a quadratic term,~$\delta^2$, and a higher derivative term,~$\d^2\delta$. After the inclusion of these three terms, the EFTofLSS at two-loop order matches simulation data  up to $k\simeq 0.34\hinvMpc$ at redshift~$z=0$, up to~$k\simeq 0.55\hinvMpc$  at $z=1$, and up to~$k\simeq 1.1\hinvMpc$ at $z=2$. At these wavenumbers, the cosmic variance of the simulation is at least as small as $10^{-3}$, providing for the first time a high precision comparison between theory and data.  The actual reach of the theory is affected by theoretical uncertainties associated to not having included higher order terms in perturbation theory, for which we provide an estimate, and by potentially overfitting the data, which we also try to address. Since in the EFTofLSS the coupling constants associated with the counterterms are unknown functions of time, we show how a simple parametrization gives a sensible description of their  time-dependence. Overall, the $k$-reach of the EFTofLSS is much larger than previous analytical techniques, showing that the amount of cosmological information amenable to high-precision analytical control might be much larger than previously believed. 

 \vspace{0.3cm}
\hrule

 \vspace{0.3cm}
 
 \newpage
\tableofcontents

\section{Introduction}

The Effective Field Theory of Large Scale Structures (EFTofLSS)~\cite{Baumann:2010tm,Carrasco:2012cv,Carrasco:2013sva,Carrasco:2013mua,Pajer:2013jj,Carroll:2013oxa,Porto:2013qua,Mercolli:2013bsa,Senatore:2014via,Angulo:2014tfa,Baldauf:2014qfa,Senatore:2014eva,Senatore:2014vja,Lewandowski:2014rca,Mirbabayi:2014zca,Foreman:2015uva,Angulo:2015eqa,McQuinn:2015tva,Assassi:2015jqa,Baldauf:2015tla} provides the analytical framework that allows one to compute the distribution of dark matter and galaxies at large distances as a perturbative expansion in powers of the overdensity. So far, the EFTofLSS has been compared to simulation data for the case of the dark matter density power spectrum~\cite{Carrasco:2012cv,Carrasco:2013mua,Senatore:2014via} and bispectrum~\cite{Angulo:2014tfa,Baldauf:2014qfa}, the dark matter momentum power spectrum~\cite{Senatore:2014via},  the dark-matter vorticity slope~\cite{Carrasco:2013mua,Hahn:2014lca}, the baryon power spectrum~\cite{Carrasco:2013mua}, the halo power spectrum and bispectra (including all cross correlations with the dark matter field)~\cite{Senatore:2014vja,Angulo:2015eqa}, and the dark matter power spectrum in redshift space~\cite{redshift-in-prep}. The results have been very encouraging, showing that the EFTofLSS has a percent level agreement with the numerical data to a much greater wavenumber that formerly available analytic techniques (these analytic techniques are indeed incorrect if the EFTofLSS is correct).  Maybe the most amazing result was that the EFTofLSS seemed to agree within  roughly 2\% with power spectrum data from the Coyote emulator~\cite{Heitmann:2008eq}, potentially up to the relatively high wavenumber of $k\simeq 0.6\hinvMpc$~\cite{Carrasco:2013mua,Senatore:2014via}.  Very recently, while this paper was in advanced development, Ref.~\cite{Baldauf:2015tla} appeared that attempted to analyze with great precision the behavior of the dark matter displacement field in the EFTofLSS, using novel techniques that go beyond the simple power spectrum analysis. They find the EFTofLSS to fail against simulations at $1\%$ level at $k\simeq 0.2\hinvMpc$, due to the appearance of stochastic contributions, with the EFTofLSS performing much better than former techniques.

The EFTofLSS differs from former analytical techniques for two different reasons. First, the IR-resummation of infrared modes which was observed to be necessary to be treated non-perturbatively in order to have a well-defined perturbative expansion (see for example~\cite{Eisenstein:2006nk,Crocce:2007dt,Tseliakhovich:2010bj,Taruya:2012ut,Carlson:2012bu,Bernardeau:2012aq}), is done in a radically different way than in these techniques, such as RPT~\cite{Crocce:2005xy}. According to a theorem by Frieman and Scoccimarro~\cite{Scoccimarro:1995if}, generalized to the general relativistic context in~\cite{Carrasco:2013sva,Creminelli:2013mca}, the resummation of IR modes in the dark matter power spectrum should only affect the perturbative reproduction of the BAO peak, which appears in the power spectrum as oscillations  in $k$-space. {\it Therefore, according to general relativity, a correct IR-resummation of this quantity should remove the residual oscillations in $k$-space between theory and data, without changing the UV reach of the theory compared to when the IR-resummation is not performed.} This is achieved by the IR-resummation developed in the context of the EFTofLSS in~\cite{Senatore:2014via}, but, to our knowledge, not so in all formerly available techniques (including RPT~\cite{Crocce:2005xy}).
It should be stressed that the IR-resummation developed in~\cite{Senatore:2014via} differs from former approaches in the actual implementation, which now respects general relativity, not in the conceptual fact that IR modes should be resummed to correctly reproduce the BAO peak, which had already been emphasized in different contexts~\cite{Eisenstein:2006nk,Crocce:2007dt,Tseliakhovich:2010bj,Taruya:2012ut,Carlson:2012bu,Bernardeau:2012aq}. 

The second and most important difference between the EFTofLSS and other perturbative approaches is in the way short distance nonlinearities are treated. 
These other approaches, including RPT~\cite{Crocce:2005xy} or RegPT~\cite{Taruya:2012ut}, assume (as SPT does) that the short-distance modes have a vanishing stress tensor. However, this is not an innocuous assumption: it implies that short-distance physics affects long distance dynamics {\em only} through the effect of the loops originating from perturbatively solving the nonlinear fluid equations. Rather, these loops receive a non-negligible contribution from modes so short that they are not in the perturbative regime. Even though short modes are not under perturbative control, they do affect long distance physics, and therefore need to be correctly parametrized.

The EFTofLSS generalizes SPT by allowing for the most generic contribution of short modes at long distances. This results in extending the SPT equations to fluid-like equations, where the effect of short-distance modes at long distances is encoded in an infinite series of stress-tensor-like terms. The number of terms is infinite because all possible terms allowed by general relativity are introduced, in all powers of the long wavelength fields and number of derivatives. These terms are stress-tensor-like because they do not take the form that we normally have in a Navier-Stokes fluid, because the fluctuation fields include the tidal tensor of gravity, normally absent for fluids, and most importantly because the stress tensor depends on these fields in a manner which is local in space but non local in time~\cite{Carrasco:2013mua,Carroll:2013oxa}.

The expression of the effective stress tensor in a perturbative series of long-wavelength fluctuations is what makes the EFTofLSS {\it the} correct theory of the long-distance universe, in this superseding SPT. It however comes at a cost. Contrary to SPT, the EFT has in principle an infinite series of unknown parameters. However, the situation is not as tragic as it might appear at first glance.  First, each of the terms of the effective stress tensor contribute only starting at a given order in perturbation theory, so that, to make finite-order predictions, only a finite number of terms are needed. In practice, all results obtained so far for dark matter have been obtained by using only one or two of these unknown parameters~\footnote{Very explicitly, the prediction that agrees for the power spectrum at roughly percent level, at redshift zero up to $k\sim 0.25\hinvMpc$~\cite{Carrasco:2012cv} at one loop, and up to $k\sim 0.6\hinvMpc$ ~\cite{Carrasco:2013mua,Senatore:2014via} at two loops, and for the matter bispectrum at redshift zero at one loop up to $k\sim 0.3\hinvMpc$~\cite{Angulo:2014tfa} is obtained using only one and the same parameter, $c_s(z=0)$. Similarly, the prediction of the momentum power spectrum at one loop at redshift zero up to $k\sim 0.3\hinvMpc$~\cite{Senatore:2014via} is done in principle by not only using $c_s(z=0)$, but also $\dot c_s(z=0)$. In practice, $\dot c_s(z=0)$ can be inferred with the required precision at $z=0$ by using an approximate scaling symmetry of the universe, so that $\dot c_s(z=0)$ might not be considered as a free parameter. Even if one were to consider $\dot c_s(z=0)$ as an additional parameter, one should consider that with this additional parameter the EFTofLSS is able to fit the dark matter power spectrum at all redshifts~\cite{Foreman:2015uva}, and the same is expected to hold (thought it has not been verified yet) for the bispectrum and the momentum power spectrum. The prediction of the slope of the vorticity field does not require any new parameter. To similar precision, the prediction of the baryon power spectrum up to $k\simeq 0.6\hinvMpc$ at $z=0$ requires one additional parameter~\cite{Lewandowski:2014rca}.}.  Second, the parameters need to be measured in one of the following two ways.   Either one can measure them directly  by matching the predictions of the theory to  long-wavelength observations (or to simulations, which are nothing but numerical experiments). This does not mean that the theory loses all predicting power, because each of the unknown coefficients comes with a specific functional form in wavenumber-space, so that not all information is lost. Indeed, this is the way we measure the Newton constant in general relativity~\footnote{General Relativity is indeed nothing but the Effective Field Theory of a massless spin-2 particle, and has therefore an infinite number of parameters. Luckily, and at the same time unfortunately, it is very hard to measure the effect of the additional terms.}, or the way we measure $F_\pi$ in the Chiral Lagrangian that describes pion interactions. The second way in which the parameters of the EFTofLSS can be measured is by measuring the effective stress tensor directly from dark matter particles, which, in contrast to the fluid elements, represent the correct degrees of freedom at short distances. This measurement can be done with small simulations, which only have to reproduce the nonlinear scale, and are therefore very fast and potentially more accurate. This method of measuring the parameters of the EFTofLSS leaves no free parameter when the theory is compared to data, and was pioneered in~\cite{Carrasco:2012cv}.

The reach of the EFTofLSS depends on the precision of the measurement of the paramaters, and therefore on the quality of the data available.  Previous results were obtained from emulators such as those provided by CAMB~\cite{Lewis:1999bs} or Coyote~\cite{Heitmann:2008eq}, which have at least one-percent error, or with high cosmic variance simulations.
These results seemed to show that one could could fit the data up to  $k\simeq0.6\hinvMpc$ staying within the error bars, but with an uncertainty that could push back the $k$-reach as low as  $k\simeq0.4\hinvMpc$ (see e.g.~\cite{Carrasco:2013mua} or~\cite{Senatore:2014via}). Furthermore, the objective of the first papers on the EFTofLSS was to focus on understanding the theoretical aspects of the theory, rather then to establish the precise $k$-reach of a given fixed order calculation.

In this paper,  we compare the predictions of the EFTofLSS with data from a simulation (Dark Sky~\cite{Skillman:2014qca}, described in Sec.~\ref{sec:darksky}) with very small cosmic variance down to relatively low wavenumbers. Furthermore, since we have access to more detailed, quasi-direct, information about the data, it is possible to control many of the systematic errors that can occur in the comparison between simulations and theory~\footnote{One of the authors would like to stress that it is a well known fact among people dealing with numerical (and experimental) data that to perform comparisons with exquisite precision, it is necessary to have comparable exquisite knowledge of the source of the data themselves. When previously existing data are subjected to new analyses, it can sometimes happen that systematic effects in the data that were previously irrelevant will become relevant to the new analyses, and we expect that data from $N$-body simulations are no exception.}. We use a fitting procedure that properly incorporates the cosmic variance uncertainty on the power spectrum, without trying to account for unknown systematics, finding that the prediction of the EFT at two loops, including only the lowest-order counterterm (associated with a single free parameter), agrees with the nonlinear measurements at redshift $z=0$ up to $k\simeq 0.15\hinvMpc$ to within $\sim$0.3\% (the cosmic variance errorbar at that wavenumber). This represents a reduction in the $k$-reach of the EFT with respect to former results, albeit with a much higher level of precision and accuracy of the numerical data. Note, in fact, that on the same numerical data, linear theory and two-loop SPT display a similar reduction in their $k$-reach, failing at $k\simeq 0.04\hinvMpc$ already by $1\%$, against the common knowledge that states that linear theory fails at $0.1\hinvMpc$. Indeed, calculations in the EFT represent an expansion roughly in powers of $k/\knl$, so that a calculation at a given order will fail at lower and lower wavenumber as we increase the precision of the data more and more. Furthermore, as we describe in detail in the bulk of the paper, this reduction in the $k$-reach is mainly due to a shift on the value of the speed of sound $\co$, induced in turn by the more precise numerical data. As clearly highlighted in~\cite{Foreman:2015uva}, the $k$-reach of the theory obtained in former studies was indeed highly sensitive to the numerical value of $\co$.

In previous work, we justified neglecting all other counterterms in the two-loop EFT prediction on the basis of order-of-magnitude estimates of their sizes. In this paper, we introduce a more detailed method for determining the importance of these counterterms, based on the fact that these terms should compensate for the contribution in loop corrections by modes with wavenumbers larger than the nonlinear scale. We find that the  counterterms that have previously been included are not sufficient to make the calculation UV-insensitive (that is, to correct for the bulk of the UV contribution from the loop integrals), and that two more counterterms are necessary and sufficient for this purpose. With a total of three parameters, the EFTofLSS at two loops becomes UV-insensitive, and the $k$-reach of the theory is increased to $k\simeq 0.34\hinvMpc$, where the cosmic variance of the simulation is about~$10^{-3}$.

We then investigate the impact of various other sets of counterterms, including a stochastic term, on the power spectrum prediction and its performance with respect to simulation data. We also extend the analysis to higher redshifts, finding similar results for the $k$-reach, and study the time-dependence of the counterterms.
Given that the number of modes scales as the maximum wavenumber cubed, the gain from using the EFTofLSS with respect to linear theory is still very large, somewhere between two and three orders of magnitude before considering the loss of information due to the marginalization over the three counterterms. Such an increase in the number of available modes might have huge consequences for our capabilities to explore the early universe in the next decade.

\section{Simulations and Fitting Procedure \label{sec:simulation}}

\subsection{The Dark Sky simulations}
\label{sec:darksky}

In this paper, we will compare the EFTofLSS prediction for the matter power spectrum to one of the Dark Sky series of dark matter-only N-body simulations~\footnote{\href{http://darksky.slac.stanford.edu}{http://darksky.slac.stanford.edu}}. These simulations use a Planck-like cosmology with cosmological parameters $\{ \Omega_{\rm m} , \Omega_{\rm b} , \Omega_\Lambda, h, n_{\rm s}, \sigma_8 \} = \{ 0.295, 0.0468, 0.705, 0.688, 0.9676, 0.835 \}$. Their initial conditions were set up using 2LPT, with an initial power spectrum generated by the CLASS code~\cite{Lesgourgues:2011re}.

In particular, we will utilize the \texttt{ds14\_a} run of Dark Sky, which evolved $10240^3$ particles in a box of side length $8 h^{-1} {\rm Gpc}$ from an initial redshift $z_{\rm init}=93$ to $z=0$. The matter power spectrum was measured from \texttt{ds14\_a} snapshots at various redshifts on a Fourier grid with $8192^3$ cells, with a constant shot noise contribution subtracted analytically, and averaged into bins of width $\Delta k= 2\pi/L_{\rm box} \approx 8\times 10^{-4} \invMpc$. The cosmic variance errorbar on the value of each bin is estimated by the standard Gaussian formula,
\beq
\frac{\delta P_i}{P_i} = \sqrt{ \frac{2}{N_i} }\ ,
\eeq
where $N_i$ is the number of modes contained in bin $i$. Where this number falls below $10^{-3}$, we instead use $\delta P_i/P_i = 10^{-3}$ as the errorbar, because this is roughly the precision of the theoretical computations that we will compare to the data.

The combination of high resolution and large box size suppresses the cosmic variance of the power spectrum to sub-percent levels for $k\gtrsim 0.05\invMpc$, allowing a detailed comparison between theory and nonlinear measurements at much smaller wavenumbers than previously possible. Note, however, that this cosmic variance is already present in the (linear) initial conditions, rather than being significantly altered by nonlinear evolution. Therefore, when plotting various power spectrum predictions we replace the theoretical, averaged linear spectrum (generated by CLASS) by the spectrum measured from the initial conditions of \texttt{ds14\_a} (rescaled to the appropriate redshift). We make  this replacement only for $k<0.1\invMpc$. By doing so, we can dramatically reduce the low-$k$ variance between the predictions and numerical data, as seen in the ratios we will plot later in this paper.

We only make this replacement when generating the plots; the actual fits to the data (described in the next section) are performed using the theoretical linear spectrum. This is because it is more difficult to estimate the residual errorbars on the simulation measurements due to cosmic variance if the measured $P_{11}$ is used in the theory prediction (or used to compute the loop corrections, which is also possible in principle), and the cosmic variance of \texttt{ds14\_a} is small enough that we can obtain satisfactory results without this extra complication (it would be interesting to pursue this approach in future work).

\subsection{Fitting procedure\label{sec:fittingprocedure}}

The EFTofLSS power spectrum consists of the usual SPT expansion plus a set of counterterms coming from the expansion and evaluation in perturbation theory of the dark matter effective stress tensor (on large scales).  The parameter multiplying each counterterm is unknown and is fixed by the fitting procedure, but there are two complications that arise at this point. First, there is some ambiguity about which counterterms to include in a given prediction, and this ambiguity is only partially reduced by attempting to estimate the relative size of each term, since these estimates are only accurate at the order-of-magnitude level. In Sec.~\ref{sec:UVsensitivity}, we will describe a method to approximately estimate which counterterms should be included in a given calculation, while Sec.~\ref{sec:othercounterterms} will be devoted to an exploration of various combinations of counterterms that can be included, such as a stochastic term $\propto k^4$. 

The second complication is that, apart from the aforementioned order-of-magnitude estimates, we do not know the maximum reach of the theory a priori.  However, we can use the following fact to our advantage: as one increases $\kmax$ (the maximum wavenumber used for a fit), we observe that the parameters change by remaining within the error bars of the fit (which shrink as we move to higher $\kmax$, because as we increase $\kmax$ we use more data points), and beyond some $\kmax$, the parameters start changing beyond the amount expected from the error bars. We interpret this as the fact that, by adding data, the parameters are better constrained and converge to their actual values, staying approximately constant as a function of $\kmax$. When $\kmax$ enters the region where higher order terms become relevant, the best fit parameters change in a statistically unexpected way because the fitting method tries to compensate for the omitted terms.  
We therefore define $\kfit$ as the maximum~$\kmax$ for which the parameters have stable values and $\kreach$ to be the corresponding reach of the theory for these values of the parameters, that is, the maximum $k$ for which the theory prediction is within the error bars of the data.\footnote{Since our goal in this paper is to perform a controlled comparison between the EFTofLSS and the Dark Sky simulations, we find it natural to use the uncertainty on the simulation output to define when the theory ``fails." Instead, if one is comparing to another simulation or to observational data, a different definition of ``failure" could certainly be more appropriate. Note, however, that near-term observations are unlikely to require a precision as stringent as we require in this work, so in this light, our assessment of the failure of the power spectrum prediction can be seen as somewhat pessimistic.} Note that $\kreach$ is not necessarily equal to $\kfit$. In principle, the theory curve could continue to match the data beyond $\kfit$. In practice, this does not happen because, by construction,~$\kreach$ is the maximum $\kmax$ beyond which the values of the parameters that would match the data change significantly. For the convenience of the reader, we summarize the definitions of $\kmax$, $\kfit$, and $\kreach$ in Table~\ref{tab:kdefs}.

   \begin{table}[t]
\begin{center}
   \begin{tabular}{ c | l}
Notation & Definition \\
\hline
$\kmax$ & Maximum $k$ of power spectrum measurements used in a fit \\ 
$\kfit$ & For a given prediction, the maximum value of $\kmax$ for which the \\
& \quad fit parameters are ``stable" (as defined in the main text) \\ 
$\kreach$ & The wavenumber at which the prediction fails, \\
& \quad when parameters are fit using the region $k<\kfit$ \\ 
\end{tabular}
\caption{\label{tab:kdefs} \small   Descriptions of various wavenumbers defined in Sec.~\ref{sec:fittingprocedure} and used throughout the text.}
   \end{center}
   \end{table}

The stable region of the parameters can be identified by determining the $2\sigma$ region associated to the fit of each parameter at a given $\kmax$. As we increase $\kmax$, we expect the determination of the same parameter at the higher $\kmax$ to lie within the $2\sigma$ region of the lower $\kmax$, as otherwise the fit obtained with the new parameter up to the smaller $\kmax$ would be significantly worse~\footnote{The criterion for allowing for a $2\sigma$ discrepancy is a bit arbitrary. We could have chosen, for example, a $1\sigma$ discrepancy, even though we would find a bit odd to forbid fluctuations in the parameters of more than $1\sigma$. Our estimate of the theoretical error will nevertheless be such that it will encompass the case if we had chosen a $1\sigma$ threshold. }. We present figures of the parameters as a function of $\kmax$ later in the text, and we use this as a criterion to determine $\kfit$. We also check that the stable region is always within a range of wavenumbers where the $p$-value  of the comparison between theory and data is close to one. We believe that this method should help prevent overfitting. In the next section, each plotted power spectra corresponds to the best fit value of the parameters for the $\kfit$ determined by this procedure, even though we could practically chose any $\kfit$ within the stable region, as the theory curve would not change much if the parameters were taken with a $\kfit$ everywhere in the stable region. 

 We estimate the theory error by plotting the values obtained by choosing one of the parameters to be $1\sigma$ away from the best fit value that we have at $0.75\kfit$, and then re-fitting for the other free parameters in the prediction. Since physical results should be independent of the renormalization scale, this procedure encapsulates the effect of higher-order terms that are not included. This procedure is also affected by the smaller amount of data that we have at a lower $\kfit$, which affects the determination of the parameters. Unfortunately, in an EFT, estimates of any results, such as the $k$-reach or the theoretical error, are most reliably performed using the data themselves. One could use purely theoretical estimates such as those provided in~\cite{Foreman:2015uva}, but they agree with the ones used here, given the fact that the theoretical error should be taken at the order of magnitude level.    

For numerical computations of the loop corrections in the EFTofLSS, we use a modified version of the Copter code~\cite{Carlson:2009it} that makes use of the IR-safe integrands of~\cite{Carrasco:2013sva} and the Monte Carlo integration capabilities of the CUBA library~\cite{CUBA}. We use the same input linear power spectrum as the Dark Sky simulations, and require a relative numerical precision of  $10^{-3}$ on all results. The IR-resummation method of~\cite{Senatore:2014via} is then implemented in Mathematica, and is used for all EFTofLSS predictions presented in this paper.

\section{The Two-Loop Power Spectrum}

In this section, we discuss the two-loop matter power spectrum in the EFTofLSS. We first review the formulas that have previously appeared in the literature, describe procedures for determining the free parameters $\co$ and $\ct$ associated with them, and present a comparison to redshift-zero measurements from the Dark Sky simulation. We then move on to re-examine the choice of which counterterms to include in the calculation, and find that considerations of UV-sensitivity necessitate the inclusion of two additional counterterms. The final subsection compares this three-counterterm prediction to data.

In the conclusions of this paper, we will comment on possible numerical problems that might affect the precise comparison of the EFTofLSS with simulations. For the moment, we will proceed without questioning the reliability of the comparison we are performing.

\subsection{Formulas and renormalization}
\label{sec:formulas}

 We will begin by reviewing the one- and two-loop EFTofLSS predictions for the power spectrum as they have appeared previously in the literature. The one-loop EFT formula for the power spectrum is given by
\beq
P_\text{EFT-1-loop}(k,z) = [D_1(z)]^2 P_{11}(k) + [D_1(z)]^4 \poneloop(k) + \ptreecs(k,z)\ ,
\label{eq:peft1loop}
\eeq
where
\beq
\ptreecs(k,z) = -2(2\pi) \co(z) [D_1(z)]^2 \frac{k^2}{\knl^2} P_{11}(k)\ ,
\eeq
while the two-loop formula is
\begin{align} \nn
&P_\text{EFT-2-loop}(k,z) = P_\text{EFT-1-loop}(k,z) + [D_1(z)]^6 \ptwoloop(k)
	-2(2\pi) \ct(z) \frac{k^2}{\knl^2} P_{11}(k) \\
&\quad + (2\pi) \co(z) [D_1(z)]^4 \poneloopcs(k) 
	+ (2\pi)^2 \lp 1 + \frac{\zeta+\frac{5}{2}}{2(\zeta+\frac{5}{4})} \rp [\co(z)]^2 [D_1(z)]^2 \frac{k^4}{\knl^4} P_{11}(k)\ .
\label{eq:peft2loop}
\end{align}
Expressions for $\poneloop(k)$ and $\ptwoloop(k)$ are given in~\cite{Carrasco:2013sva,Carrasco:2013mua}. Expressions for $\poneloopcs(k)$ are given in~\cite{Carrasco:2013mua} for the $\zeta=2$ case; the extension to $\zeta\neq 2$ is straightforward using the recurrence relations from~\cite{Angulo:2014tfa}. The derivation of the $(k/\knl)^4 P_{11}$ term, along with a discussion of the $\zeta$ parameter and how it is related to the time-dependence of the counterterms, can be found in~\cite{Foreman:2015uva}.

Implementation of these formulas in a comparison against numerical data requires the determination of the parameters $\co$ and $\ct$. The procedure for doing so was essentially laid down in~\cite{Carrasco:2013mua}, modulo a few changes that we will highlight.  If we only wish to use the one-loop prediction, Eq.~(\ref{eq:peft1loop}), we only need the value of $\co$, which can be found from the procedure described in Sec.~\ref{sec:fittingprocedure}: choosing the best fit value of $\co$ up to the $\kmax$ for which it is relatively constant and the $p$-value is acceptable. One can also determine $\co$ directly from the two-loop prediction~(\ref{eq:peft2loop}), after $\ct$ has been fixed in a way that we will describe below. This is the main procedure we will use in this work, for two reasons. First, the two-loop prediction will match the data over a larger region than the one-loop prediction, enabling a more precise determination of $\co$. Second, there is a larger risk of overfitting, and therefore of obtaining an incorrect value of $\co$, if only the one-loop prediction is used (we comment further on this in Sec.~\ref{sec:z0comp}.)

\begin{figure}[t]
\begin{center}
\includegraphics[scale=0.7]{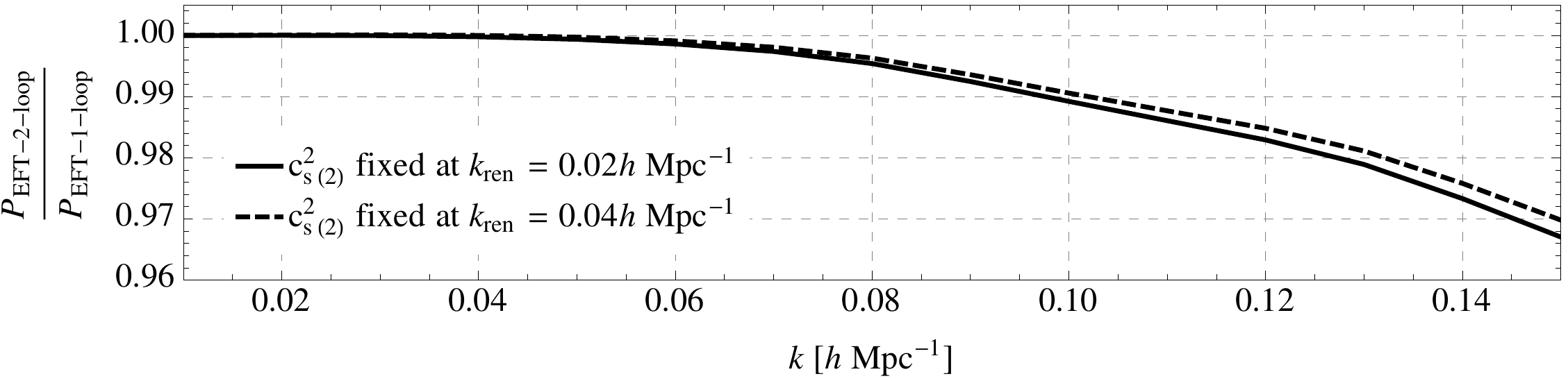}
\caption{ \small Ratio of the predictions of the EFT at two and one loops. By choosing a value of $\kren$ small enough, we are able to make this ratio very flat, demonstrating that the higher order terms contained in $P_\text{EFT-2-loop}$ are negligible below at the wavenumber at which we impose the renormalization condition. }\label{fig:ratio2to1}
\end{center}
\end{figure}

We now describe how to determine $\ct$. As discussed in~\cite{Carrasco:2013mua}, when we evaluate the two-loop diagrams, we find that, when the wavenumbers running in the loops are taken to be much larger than the external wavenumber $k$, these loops produce a contribution that is functionally of the form $k^2P_{11}(k)$. This contribution is degenerate with the lowest-order counterterm $\ptreecs$, associated with $\co$. The simplest way to handle this degeneracy is to choose $\ct$ in order to cancel the part of the two-loop terms that scales like $k^2 P_{11}$.  To do so, we can use the fact that the term $k^2P_{11}(k)\subset \ptwoloop$ is the one that dominates as $k\to 0$~\footnote{Locality in space, as well as momentum and matter conservation, forbid any term that is generated by modes higher than $k$ to decay more slowly than $k^2P_{11}(k)$ at low $k$'s.}. Therefore, we can determine $\ct$ by choosing a very low wavenumber $\kren$, at which the terms that decay faster than  $k^2P_{11}(k)$ have become negligible, and require that
\be\label{eq:renormalization}
P_\text{EFT-1-loop}(\kren,z)=P_\text{EFT-2-loop}(\kren,z) \qquad\Rightarrow\qquad \ct=\ct(\kren, \co)\ .
\ee
Any sizeable non-logarithmic dependence of $\ct$ on $\kren$ should be taken as an indication that the renormalization scale has not been taken infrared enough for all the terms in $\ptwoloop$ other than $k^2P_{11}(k)$ to be negligible. If $\kren$ is taken sufficiently low that logarithmic running of $\ct$ is not expected (based on the slope of the linear power spectrum), then $\ct$ should be essentially independent of $\kren$.

The determination of $\ct$ can be made easier by a manipulation of the two-loop integral. Since any term that is of the form of $k^2 P_{11}(k)$ is irrelevant to compute, as it amounts to a redefinition of $\ct$, we define a ``UV-improved" version of $\ptwoloop$ by subtracting from $\ptwoloop$ a term which is a good approximation of its $k^2 P_{11}(k)$ component:
\be
\ptwoloop^\text{(UV-improved)}(k)
= \ptwoloop(k)- 2\, k^2 P_{11}(k)\,
\int_{\{q_1,q_2\}> k_{\rm min}}^\infty \frac{d^3q_1}{(2\pi)^3}\,\frac{d^3q_2}{(2\pi)^3}\; 
\lim_{\tilde k\to 0} 
\frac{P_{51}^{\rm integrand}(\tilde k,\vq_1,-\vq_1,\vq_2,-\vq_2)}{2 \tilde k^2 P_{11}(\tilde k)}\ ,
\ee
where $P_{51}^{\rm integrand}(k,\vq_1,-\vq_1,\vq_2,-\vq_2)$ is such that the usual SPT diagram $P_{51}(k)$ is equal to 
\be
P_{51}(k)
= \int_{k_{\rm min}}^\infty \frac{d^3q_1}{(2\pi)^3}\,\frac{d^3q_2}{(2\pi)^3}\; 
P_{51}^{\rm integrand}(k,\vq_1,-\vq_1,\vq_2,-\vq_2)\ .
\ee
Using $\ptwoloop^\text{(UV-improved)}$ has the advantage that, if $\kren$ and $k_{\rm min}$ are taken low enough, the $\ct$ that will be obtained from Eq.~(\ref{eq:renormalization}) will be very small, and in fact vanishingly small apart from running effects~\footnote{%
$\ptwoloop^\text{(UV-improved)}$ also has a significant advantage in terms of computational cost, because the UV-sensitive term $k^2P_{11}(k)$ that is removed from $\ptwoloop$ is typically much larger than the UV-insensitive term, $\ptwoloop^\text{(finite)}$, that we are interested in calculating, in the range of wavenumbers relevant for this work. Even in $\ptwoloop^\text{(UV-improved)}$, there will be some residual UV-sensitive terms, but these are also much smaller than $\ptwoloop^\text{(finite)}$. Therefore, for a given level of accuracy that we desire for $\ptwoloop^\text{(finite)}$, a smaller number of evaluations of the integrand (around one order of magnitude fewer, in fact) is required when computing $\ptwoloop^\text{(UV-improved)}$. The algebraic expression for $\ptwoloop^\text{(UV-improved)}$ is much more complicated than for $\ptwoloop$, causing the UV-improved integrand to be $\sim$10-20\% slower to evaluate than the regular integrand, but this does not interfere with the significant computational gains of $\ptwoloop^\text{(UV-improved)}$.
%
 }. 

From this discussion, it is clear that, once $\ct$ is determined from (\ref{eq:renormalization}), the ratio $P_\text{EFT-2-loop}/P_\text{EFT-1-loop}$ remains close to one up to a higher $k$ than if a different value of~$\ct$ was to be chosen, as this ratio goes as $1+\ptwoloop^\text{(finite)}/P_\text{EFT-1-loop}$, which is much closer to one than $1+(k/\knl)^2P_{11}(k)/P_\text{EFT-1-loop}$. Fig.~\ref{fig:ratio2to1} presents the ratio $P_\text{EFT-2-loop}/P_\text{EFT-1-loop}$ for two different values of $\ct$ obtained by applying Eq.~(\ref{eq:renormalization}) at two different renormalization scales. The ratio has the expected behavior: it is quite close to one up to $k\simeq 0.1\hinvMpc$, where $\ptwoloop^\text{(finite)}$ causes it to deviate from one by $\sim 1\%$. Indeed, this is  the scale at which the one-loop EFT fails at one-percent level, as it lacks $\ptwoloop^\text{(finite)}$; furthermore, we have checked that as soon as we  relevantly change the value of $\ct$, the ratio deviates from one at a much lower $k$. Finally, we find that the ratio shows the same behavior whether we use the UV-improved version of $\ptwoloop$ or not; in the former case, we find $\ct$ of order $0.02\left(\knl / (2 \invMpc) \right)^2$, while in the latter case we find $\ct \sim -2\left(\knl / (2 \invMpc) \right)^2$. The difference between these two values is the contribution to $\ct$ that is removed by the UV-improved procedure. For the rest of this work, we will use the UV-improved version of $\ptwoloop$, $\ptwoloop^\text{(UV-improved)}$, and choose $\ct=0$. This procedure can be thought of as choosing $\kren\simeq k_{\rm min}$, and we will take $k_{\rm min}\simeq  5\times 10^{-4} \invMpc$, which is much lower than the lowest wavenumber we will use in our comparison between theory and data.

\subsection{Comparison to data at $z=0$}
\label{sec:z0comp}

\begin{figure}[t!]
\begin{center}
\includegraphics[scale=0.75]{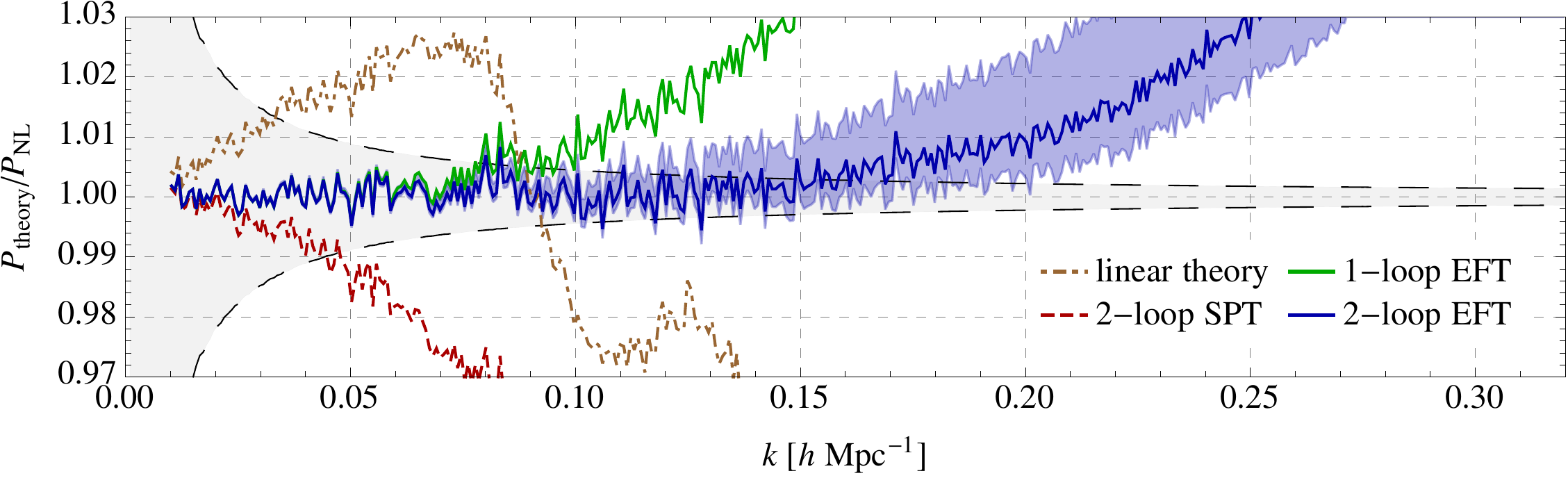}
\caption{ \small  The two-loop EFTofLSS prediction for the $z=0$ power spectrum, when one includes only one counterterm (associated with the speed of sound $\co$), along with various other theory predictions. The EFT curves use a value of $\co \simeq 0.53\left(\knl / (2 \invMpc) \right)^2$. We can see that the theory performs better and better as higher order contributions are included. The blue shading represents the variation of the result if we perform the fit to determine $\co$ up to $0.75\kfit$ and choose the two values $1\sigma$ away from the central value, where $\kfit$ is the wavenumber beyond which $\co$ begins to deviate from the value determined at lower $k$.  For $k<0.1\invMpc$, the linear power spectrum in the theory prediction is replaced with the power spectrum measured from the initial conditions of the simulations, allowing for a dramatic reduction in the variance of the $P_\text{theory}/P_\text{NL}$ curves at these wavenumbers, but also implying that the cosmic variance errorbars (represented by the grey shading) do not reflect the uncertainty on the curves for $k<0.1\invMpc$ (as discussed in Sec.~\ref{sec:darksky}). The $\kreach$ of the EFT at two loops is about $\kreach\simeq 0.15\hinvMpc$, where the cosmic variance is about 0.4\%, even though there is large theoretical uncertainty. The $k$-reach is smaller than what was previously presented in~\cite{Carrasco:2013mua}, where the errorbars were taken to be $\sim$2\%, because the much higher precision of the available numerical data allows the choice of a lower $\kren$, which eliminates the strong cancellation between various two-loop terms that was seen in~\cite{Carrasco:2013mua}.  }\label{fig:two-loop-prediction}
\end{center}
\end{figure}

\begin{figure}[t!]
\begin{center}
\includegraphics[scale=0.75]{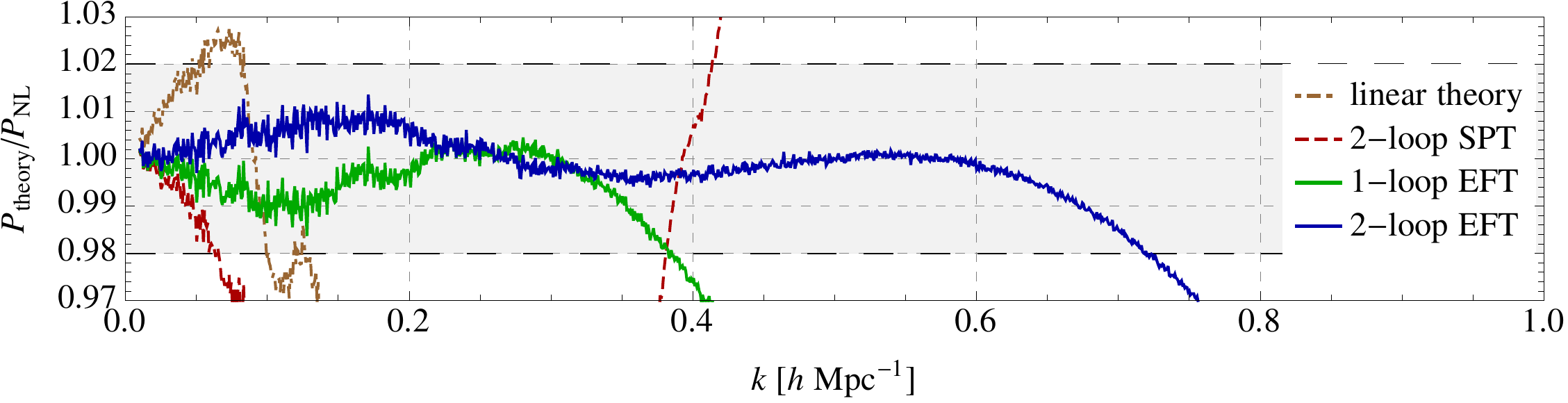}
\caption{ \small   Same as Fig.~\ref{fig:two-loop-prediction}, but using the procedure from previous papers (e.g.~\cite{Carrasco:2013mua,Senatore:2014via,Foreman:2015uva}) to fix $\co$ and $\ct$ (see the main text for details). If we allow for a uniform 2\% error budget on $P_{\rm NL}$ (as was required in previous work, due to the use of the Coyote emulator), and use a $\kren$ comparable to previous work ($\kren\sim 0.23\invMpc$), we find results that are consistent with~\cite{Carrasco:2013mua,Senatore:2014via,Foreman:2015uva} (the $\sim$1.5\% offset between the one- and two-loop curves was actually about $0.7\%$ in the cosmologies and data considered in~\cite{Carrasco:2013mua,Senatore:2014via,Foreman:2015uva}). One sees that at low $k$, there is an offset between theory and data. Such an offset cannot be ruled out by using power spectra from the Coyote emulator, since its output has a systematic errorbar of at least $1\%$. Instead, for Dark Sky, we cannot allow for that level of systematic error, and forcing the absence of the offset at low $k$ affects the $k$-reach of the theory, leading to the results in Fig.~\ref{fig:two-loop-prediction}.}
\label{fig:two-loop-prediction-high-kren}
\end{center}
\end{figure}

Before re-examining the self-consistency of the two-loop prediction in Eq.~(\ref{eq:peft2loop}), we will first compare it to the matter power spectrum measured from the Dark Sky simulation at redshift $z=0$, in order to enable an easier comparison with previous results.  From Fig.~\ref{fig:two-loop-prediction}, we can see that the UV reach increases as each set of higher-order terms is added: the linear prediction fails at $\kreach\simeq0.035\hinvMpc$ by $\sim1.4\%$,  at one loop at $\kreach\simeq0.08\hinvMpc$ by $\sim 0.5\%$, and at two loops at $\kreach\simeq0.15 \hinvMpc$ by $\sim 0.4\%$. In contrast, we see that there is no relevant improvement between linear theory and two-loop SPT (and all analytic techniques prior to the EFTofLSS such as RPT and RegPT, which differ from SPT only by the resummation of the IR-modes, which are irrelevant for the UV reach of the theory~\footnote{This point is quite unappreciated in the literature, so we repeat it here. RPT is sometimes  claimed to improve the UV reach of the theory. However, to our understanding, it is supposed to be just an IR-resummation, and therefore if it improves the broad-band UV reach of the theory it violates General Relativity. It is therefore incorrect and should not be considered {as a way to increase the broad-band UV reach}. Alternatively, one should consider RPT as a fitting function. See~\cite{Lewandowski:2014rca} for a more detailed discussion. It should not be forgotten that the fact that RPT violates General Relativity (and cannot therefore be a correctly implemented IR-resummation) was already pointed out in the original RPT paper~\cite{Crocce:2007dt}, whose  focus indeed was not on the broad-band $k$-reach.}).  The blue shaded region in Fig.~\ref{fig:two-loop-prediction} represents the difference between fitting up to $\kfit$ or instead fitting up to $0.75\,\kfit$ and choosing the values of $\co$ $1\sigma$ away from the best fitting point; this represents a rough estimate of the theoretical error associated with the prediction of the EFTofLSS at this order, and should be taken at the order of magnitude level.

We now make a few comments on these results. The first is on the importance of the IR-resummation. Without performing the IR-resummation, the EFT prediction would be off with respect to the data by oscillations of order $\sim 2\%$ (see~\cite{Carrasco:2013mua}). Due to the smallness of the current error bars, it would therefore be impossible to impose the theory to have a good fit to the data, and so we see that IR-resummation is essential to performing this high-precision  comparison. 

Next, we note that we have chosen a smaller~$\kren$ than in previous works, starting from~\cite{Carrasco:2013mua}, allowing for a smaller theoretical error. Notice that since the renormalization procedure to determine $\ct$ depends only on the theory calculation,  it would have been possible to choose a low $\kren$ also when comparing with more noisy data, such as the output of the Coyote emulator. This procedure is correct, but it has the technical inconvenience that, as we explain in the next paragraph, one would have not been able to compare the value of $\co$ at one loop and at two loops, which is an interesting, though delicate, consistency check of the theory, that was performed in~\cite{Carrasco:2013mua}. This procedure is also  more sensitive to  the evaluation of the numerical loops and IR-resummation, as the overall effect of $k^2P_{11}$ drops rapidly. 

Overall, it would be incorrect to interpret the higher $k$-reach of~\cite{Carrasco:2013mua} as simply due to a choice of too large a value of $\kren$.  The renormalization scale chosen in~\cite{Carrasco:2013mua} is actually not particularly high, when we consider that the low-$k$ offset between $P_\text{EFT-1-loop}$ and  $P_\text{EFT-2-loop}$ at the $\kren$ chosen in~\cite{Carrasco:2013mua} is less than $0.5\%$, which is not large given the error bars on the data in~\cite{Carrasco:2013mua}. The reason why this choice leads to a sizable effect on the~$\kreach$ of the theory is due to the fact that, as explained in detail in~\cite{Foreman:2015uva}, the value of $\co$ that happens to be chosen in this procedure leads to an accidental cancellation between $\ptwoloop^\text{(finite)}$ and $\poneloopcs$. When combined with the poor precision of the numerical data used in~\cite{Carrasco:2013mua}, the $\kreach$ is made larger. For reference, in Fig.~\ref{fig:two-loop-prediction-high-kren} we show what happens when we perform fits to the Dark Sky data using the procedure from previous papers~\cite{Carrasco:2013mua,Senatore:2014via,Foreman:2015uva}, also allowing for a larger error budget on the nonlinear power spectrum.

We also wish to highlight the importance of the two-loop contribution in determining $\co$ from the power spectrum. Without the inclusion of the two-loop terms, one could choose the value of $\co$ in such a way as to potentially match the data up to $k\simeq0.11\hinvMpc$. If $\co$ is instead determined from the full two-loop prediction, the value of $\co$ we find, $\co \simeq 0.53 \lp \knl / (2 \invMpc) \rp^2$, is roughly~30\% smaller than that obtained from the one-loop fit, $\co \simeq 0.75 \lp \knl / (2 \invMpc) \rp^2$, and causes the one-loop prediction to fail at $k\simeq0.08\hinvMpc$. This reveals that using the one-loop prediction on its own potentially leads to overfitting.  
 
Indeed, in the limit of infinitely precise data, we would like to determine $\co$ at as low a wavenumber as possible. In practice, though, there is a minimum wavenumber below which the effect of the leading counterterm becomes smaller than the uncertainty on the data, and this wavenumber will act as a lower bound for the region where it safe to fit for $\co$. This minimum wavenumber is around $k\sim 0.06 \invMpc$ (estimated by when $P_\text{tree}^{c_s}/P_{11}$ equals the uncertainty on our data). Meanwhile, the one-loop prediction begins to fail around $k\sim 0.08 \invMpc$, but fitting up to this point would introduce a substantial bias on the resulting value of $\co$, since, by our definition of ``failure," at $k\sim 0.08 \invMpc$  the two-loop terms contribute by an amount equal to the uncertainty on the data, which in turn is similar to the size of the effect of $\co$ at these wavenumbers. Therefore, the region in which it is safe to fit for $\co$ using $P_\text{EFT-1-loop}$ alone cannot be larger than roughly $0.06 \invMpc \lesssim k \lesssim 0.07 \invMpc$, but it would be impractical to try to fit over such a small region. On the other hand, our procedure of fitting for $\co$ directly using the two-loop prediction allows us to use a much larger range of wavenumbers.
 
 Furthermore, as mentioned earlier in this section, another contribution to the mismatch of $\co$ from the best fit we obtain at one loop and at two loops could be associated to the fact that our two-loop renormalization procedure can be thought of as determining $\ct$ at low wavenumbers, and~$\co$ at around $\kren\sim \kreach$ at two loops. The one-loop renormalization procedure instead determines~$\co$ around the $\kreach$ of the one-loop computation. If the two-loop contribution were to have a logarithmic running at high wavenumbers, the value of $\co$ that we obtain by renormalizing the two-loop theory at high wavenumbers would be different than the one obtained by renormalizing at low wavenumbers, as it would need to absorb the running of~$\ct$ between the two renormalization scales, which we do not account for. There are also effects due to the fact that the measurement of $\co$ is done at a non-vanishingly small value of $\kren$, and that vanish only for $\kren\to 0$. Overall, this mismatch has no effect on the two-loop result present in this paper (because it  is degenerate with the additional counterterms), nor on the one-loop result as well if we allow for $\co$ to have different values at one and two loops. Since, due to the non-scale-free nature of our universe, precisely determining if the one-loop or two-loop diagrams have a logarithmic contribution is very hard if not impossible, and given the fact that this discussion is relevant only for comparing one- and two-loop results, not for the maximal reach of one- and of two-loop calculations, we do not explore this issue further.

\subsection{UV-sensitivity of the standard two-loop calculation\label{sec:UVsensitivity}}

The results of the previous section show that in order for the theory to agree with data beyond $k\simeq 0.15\hinvMpc$, we need to add more counterterms or loops to the two-loop formula given in~(\ref{eq:peft2loop}). 
Following the notation of~\cite{Angulo:2014tfa}, the next terms that are supposed to contribute to the effective stress tensor~$\gammai{}^i $ are 
\begin{itemize}
\item terms quadratic in the fields, which (as discussed in App.~\ref{app:quadterms}) take the form
\bea\label{eq:counter_quad}
\gammai{}^i &\supset\quad 
(1-\delta)\times \left\{ \d^i\left(\frac{\d_j v^j}{-\mathcal{H}(a) f}-\delta\right)\, ,\ \d^i \lb \d^2\phi \rb^2 \, ,
\  \d^i  \lb \d^j \d^k\phi \, \d_j \d_k \phi \rb\,,\  \d^i \d^j\phi\, \d_j \d^2 \phi \right\} ;
\eea
\item higher derivative terms:
\bea\label{eq:counter_high_der}
\gammai{}^i\quad \supset\quad \d^2\d^i\delta\ ;
\eea
\item stochastic terms:
\bea\label{eq:stoch2}
\gammai{}^i\quad \supset\quad \d^i\Delta\tau\ .
\eea
(See~\cite{Carrasco:2012cv} for the definition of $\Delta\tau$.)
\item cubic counterterms:
\bea
\gammai{}^i &\supset\quad   \d^i\delta^3\, ,\  \ldots \ .
\eea
Since there are many of those, we just wrote a representative one.
\end{itemize}

Recall that the two-loop formula in~(\ref{eq:peft2loop}) was derived from a form of the effective stress tensor that neglects all of the above terms. It might seem particularly strange for a two-loop calculation not to include the quadratic terms. The justification provided in~\cite{Carrasco:2013mua} was that the quadratic counterterms are included to remove the UV-sensitivity of diagrams that are not enhanced by as many factors of $(2\pi)$ as $\co$ and $\ct$ are. This reasoning is only partially correct, as shown in~\cite{Angulo:2014tfa} in the context of the bispectrum. If we focus on the one-loop bispectrum for simplicity, at the order at which we are working, only three of the terms in~(\ref{eq:counter_quad}) give rise to non-degenerate bispectrum shapes. In this case, if we take for simplicity a no-scale universe with slope $n=-1$, where the corresponding diagrams are logarithmically divergent, we find that the resulting three counterterms remove divergences that are all enhanced by one factor of $(2\pi)$, but, for two of of the three coefficients, they are suppressed by numerical factors that undo the enhancement by $(2\pi)$~\footnote{The reason why this happens, also at finite momenta, is that an integrand of the form $\int d^3q\; q_i q_j$ is not rotationally invariant, but it can be written as $\frac{\delta_{ij}}{3}\int d^3q\; q^2$. So, even though there is a factor of $(2\pi)$, the resulting enhancement is reduced by the numerical factor ($1/3$) originating from the non-rotational-invariance of the integrand. In the case of the three quadratic counterterms, one diagram is numerically enhanced to reduce such a suppression.}.  Therefore, this type of $2\pi$-counting argument turns out not to be particularly robust, and so we look for a more accurate strategy to assess which terms should be included at a given order in perturbation theory. Indeed, we can make a stronger distinction between which counterterms should be included in a two-loop calculation by examining the UV-sensitivity of the two-loop integrals directly, as we are going to explain next.

Before describing the relevant procedure, let us enumerate the counterterms in the power spectrum that result from the stress tensor terms we listed above:
\begin{itemize}
\item quadratic counterterms:
\begin{align}\label{eq:contribution_quad_counterterms}
&P_\text{quad.\ counterterms}(k,z) =\\  \nn
&\quad    \frac{(2\pi)}{\knl^2}D_1(z)^4
\left(c_0(z)\, P^\text{(quad,\,0)}_\text{1-loop}(k)+c_1(z)\, P^\text{(quad,\,1)}_\text{1-loop}(k)
+c_2(z)\, P^\text{(quad,\,2)}_\text{1-loop}(k)
+c_3(z)\, P^\text{(quad,\,3)}_\text{1-loop}(k)\right)\, ,
\end{align}
where $P^\text{(quad,\,0,1,2,3)}_\text{1-loop}(k)$ can be found using the techniques and the kernels in the Appendix of~\cite{Angulo:2014tfa}, and we have kept only the contribution that is not degenerate with $k^2P_{11}(k)$;
\item higher-derivative counterterm:
\be\label{eq:highr_contribution}
P_\text{4-deriv.\ counterterm}(k,z) = 2 (2\pi)^2 D_1(z)^2\,c_4(z) \left(\frac{k}{\knl}\right)^4 P_{11}(k)\ .
\ee
\item stochastic counterterm:
\be\label{eq:P_stoch}
P_\text{stoch}(k,z)=(2\pi)^2D_1(z)^2\,\cstoch(z)\left(\frac{k}{\knl}\right)^4\frac{1}{\knl^3} \ .
\ee
\item cubic counterterms: they are degenerate with $k^2P_{11}(k)$, and therefore do not need to be included in the calculation.

\end{itemize}
The time-dependence of each term is accounted for by the growth factor $D_1(z)$ and by the explicit time-dependence of the coefficients. The factors of $(2\pi)$ have been chosen in such a way that these terms have the same numbers of $(2\pi)$'s associated with the two-loop diagrams that generate these terms.

As previously stated, the role of the counterterms in the EFTofLSS is to make the result of a calculation insensitive to the loop integrals that are performed in perturbation theory when the integrands are evaluated at high wavenumber where perturbation theory does not apply, and to instead parametrize the correct contribution of short distance physics at large distances. This suggests that a way to check that our two-loop calculation is consistent is to ask how much it depends on the region of the two-loop integral evaluated on momenta larger than the nonlinear scale. If the result is UV-insensitive, we should obtain the same result at low $k$ if we evaluate the loop integrals with cutoff  $\Lambda=0.68\hinvMpc$  (taken to be a rough proxy for the nonlinear scale at $z=0$, which we estimate to be about $2 \,  k_{\rm reach}(z=0) \simeq 0.68 \hinvMpc$) or $\Lambda=\infty$\footnote{In practice, we use $\Lambda = 60\invMpc$ for the numerical evaluation of the integral.} by simply readjusting $\ct$. Equivalently, if we use the UV-improved integrand, we should find the same result.

\begin{figure}[t]
\begin{center}
\includegraphics[scale=0.75]{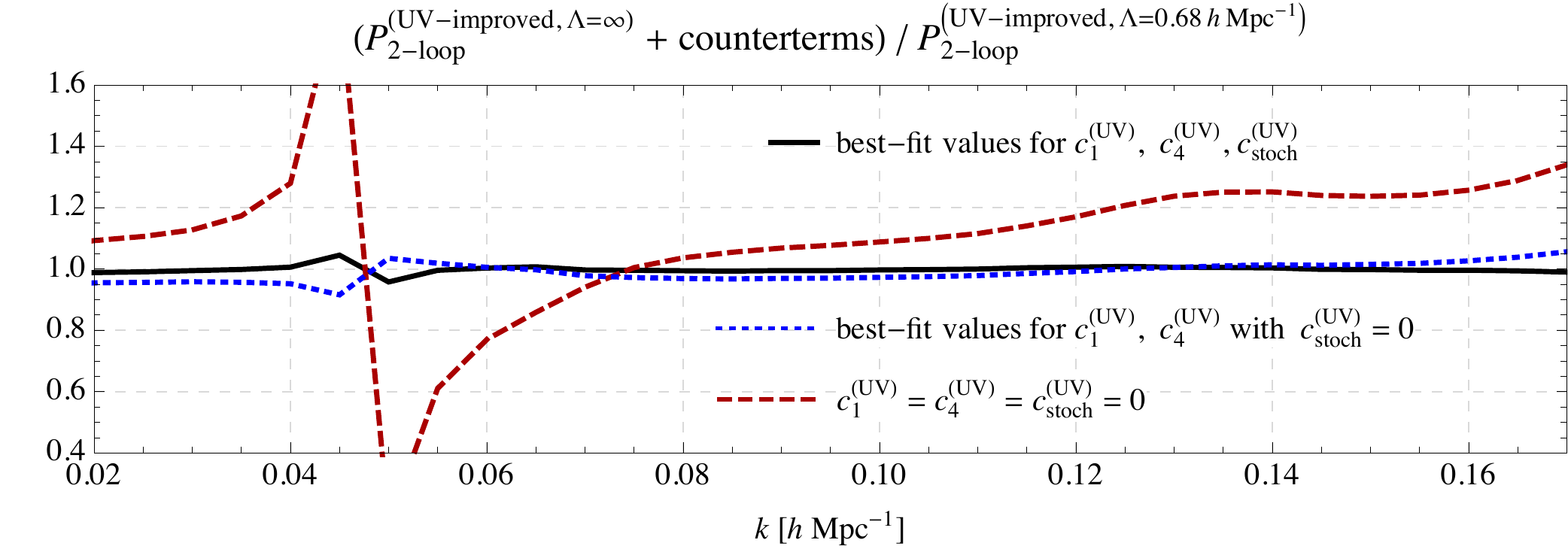}
\caption{\small The red dashed line shows the ratio of the calculations of $\ptwoloop$ with cutoff $\Lambda=\infty$ and  $\Lambda=2 \, k_{\rm reach}(z=0) =0.68\hinvMpc$, while the blue dotted and black solid lines show the same ratio but adding the two or three counterterms to the $\Lambda=\infty$ $\ptwoloop$ calculation, respectively. We see that the difference between the  $\Lambda=0.68\invMpc$  and $\Lambda=\infty$ calculation of $\ptwoloop$ can be absorbed by the counterterms. This is an important consistency check of the EFTofLSS, indicating that the unknown short-distance physics affecting the loop corrections can be accounted for by the EFT counterterms. It also tells us that the two-loop power spectrum in the EFTofLSS should minimally include the counterterms corresponding to $c_1$ and $c_4$ in order to be insensitive to our assumptions about the UV behavior of the theory.\label{fig:UV-sensitivity} }
\end{center}
\end{figure}

Upon numerically evaluating $\ptwoloop$ at these two cutoffs, we find that, to the relevant level of precision required by the simulation data, the result is indeed {\it not} the same, and that additional counterterms must necessarily be added to make the result UV-insensitive to the precision we require. In particular, we can obtain an estimate of the size of the additional counterterms that are required in order to make the result UV-insensitive by fitting the two-loop result integrated from  $\Lambda=0.68\hinvMpc$ to~$\infty$ with the additional counterterms mentioned above. In formulas, we impose
\bea\label{eq:UV-counter}
&&\left.\ptwoloop^\text{(UV-improved)}\right|_{\Lambda=0.68\hinvMpc}
- \left.\ptwoloop^\text{(UV-improved)}\right|_{\Lambda=\infty}\\
&&\quad = \frac{(2\pi)}{\knl^2}\;c_1^\text{(UV)}\; P^\text{(quad,\,1)}_\text{1-loop}(k)
+ 2(2\pi)^2c_4^\text{(UV)} \left(\frac{k}{\knl}\right)^4 P_{11}(k) 
+ (2\pi)^2\cstoch^\text{(UV)}\left(\frac{k}{\knl}\right)^4\frac{1}{\knl^3}\ . \nonumber
\eea
We add only $P^\text{(quad,\,1)}_\text{1-loop}$ and not any of the other three quadratic counterterms from~(\ref{eq:contribution_quad_counterterms}) because the shapes of these terms are quite similar.

We present in Fig.~\ref{fig:UV-sensitivity} the ratio of $\left.\ptwoloop^\text{(UV-improved)}\right|_{\Lambda=\infty}$ over $\left.\ptwoloop^\text{(UV-improved)}\right|_{\Lambda=0.68\hinvMpc}$, with the addition or not of two sets of counterterms: $\{ c_1^\text{(UV)}, c_4^\text{(UV)}, \cstoch^\text{(UV)} \}$ and $\{ c_1^\text{(UV)}, c_4^\text{(UV)} \}$. We can see that without the addition of the new counterterms, $\left.\ptwoloop^\text{(UV-improved)}\right|_{\Lambda=0.68\hinvMpc}$ is significantly different from $\left.\ptwoloop^\text{(UV-improved)}\right|_{\Lambda=\infty}$. However, after the addition of the counterterms, the result is insensitive to the contribution of the integral from $\Lambda=0.68\invMpc$ to~$\infty$, indicating that a sensible two-loop computation must minimally include the $c_1$ and $c_4$ counterterms. (Other combinations of the $c_1$, $c_4$, and $\cstoch$ counterterms, not shown, fail to absorb the UV-sensitivity to an adequate level.) The fact that that the counterterms of the EFTofLSS are able to absorb this UV-sensitivity is a nice demonstration of the internal consistency of the theory~\footnote{From a field theoretical point of view, one can argue that the correctness of the EFTofLSS is manifest already from its construction.}.

The best-fit numerical values for the coefficients $c_1^\text{(UV)}$, $c_4^\text{(UV)}$, and $\cstoch^\text{(UV)}$ give us a sense of the numerical value of the coefficients that we expect to be generated by the uncontrolled   UV physics. We find the following values when all three terms are included \footnote{If instead we take $\Lambda = 1\times k_{\rm reach}(z=0)$ we obtain 
\bea\label{eq:UV-coeff-1kreach}
&& c_1^\text{(UV)} = -2.7\left(\knl /( 2 \invMpc) \right)^2  \ , \\ \nonumber 
&& c_4^\text{(UV)} = -13\left(\knl /( 2 \invMpc) \right)^4 \ ,  \\ \nonumber 
&& \cstoch^\text{(UV)} =  13.5\times 10^4\left(\knl /( 2 \invMpc) \right)^7\ ,
\eea
and the following values when the stochastic term is omitted:
\bea\label{eq:UV-coeff-without-stoch-1kreach}
&& c_1^\text{(UV)} = -3.9\left(\knl /( 2 \invMpc) \right)^2  \ , \\ \nonumber 
&& c_4^\text{(UV)} = -17\left(\knl /( 2 \invMpc) \right)^4 \ .
\eea
These numerical values should be taken as a rough over-estimate of the size of the counterterms that is required to make the prediction UV insensitive. We clearly expect the strong coupling scale of the theory to be larger than $1\times k_{\rm reach}$. But, given our lack of knowledge of the precise value of this scale, we present values assuming that the strong coupling scale is either $1\times k_{\rm reach}$ or $2\times k_{\rm reach}$ to give a rough interval for the expected numerical values of the counterterms, an interval that we hope is large enough to include the true values.
}
\bea\label{eq:UV-coeff}
&& c_1^\text{(UV)} = -1.1\left(\knl /( 2 \invMpc) \right)^2  \ , \\ \nonumber 
&& c_4^\text{(UV)} = -5.3\left(\knl /( 2 \invMpc) \right)^4 \ ,  \\ \nonumber 
&& \cstoch^\text{(UV)} =  6.2\times 10^4\left(\knl /( 2 \invMpc) \right)^7\ ,
\eea
and the following values when the stochastic term is omitted:
\bea\label{eq:UV-coeff-without-stoch}
&& c_1^\text{(UV)} = -1.6\left(\knl /( 2 \invMpc) \right)^2  \ , \\ \nonumber 
&& c_4^\text{(UV)} = -7.0\left(\knl /( 2 \invMpc) \right)^4 \ .
\eea
The fact that the values of $c_1^\text{(UV)}$ and $c_4^\text{(UV)}$ are not strongly affected by the presence of the stochastic term is further justification that the majority of the UV-sensitivity of $\ptwoloop$ can be removed by these first two terms. Only these two counterterms need therefore to be necessarily included in order to make the two-loop calculation UV-insensitive.   

\begin{figure}[t]
\begin{center}
\includegraphics[scale=0.5]{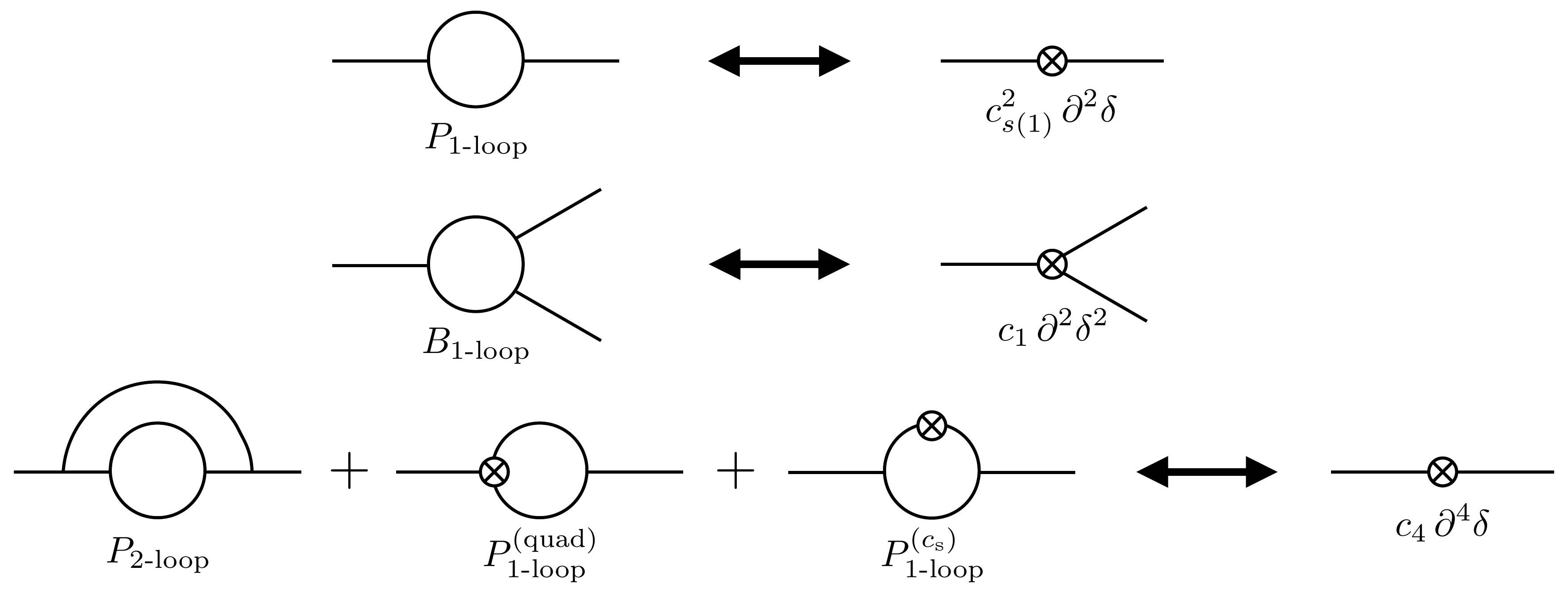}
\caption{ \small  The correspondence between the terms we add to the effective stress tensor and a schematic representation of the lowest-order diagrams (in either the power spectrum or bispectrum) they are associated with via renormalization. \label{fig:diagrams} } 
\end{center}
\end{figure}

Fig.~\ref{fig:diagrams} shows the relationship between the terms we add to the effective stress tensor and a schematic representation of the lowest-order loop diagrams (in either the power spectrum or bispectrum) they are associated with via renormalization. Note that the fact that the $\co \d^2\delta$ term renormalizes $P_{13}$, one of the terms in $\poneloop$, implies that the $c_4 \d^4\delta$ term renormalizes not just $\ptwoloop$, but also $\poneloop^\text{(quad)}$ and $\poneloopcs$, since, roughly speaking, these two terms behave similarly to $\poneloop$ but with an extra factor of $k^2$. This implies a possible physical correlation between the values of $c_4$ and $c_1$, and between $c_4$ and $\co$, but we will not explore this further in this work.

\subsection{The UV-insensitive prediction at $z=0$}
\label{sec:uvinsensitive}

As we have just argued, in order to make the two-loop calculation UV-insensitive, it is necessary to add two more counterterms: one of the quadratic terms, and the higher-derivative term. The self-consistent two-loop prediction therefore has an overall number of three parameters to be fixed from observation or simulations at a single redshift.

\begin{figure}[t]
\begin{center}
\includegraphics[scale=0.7]{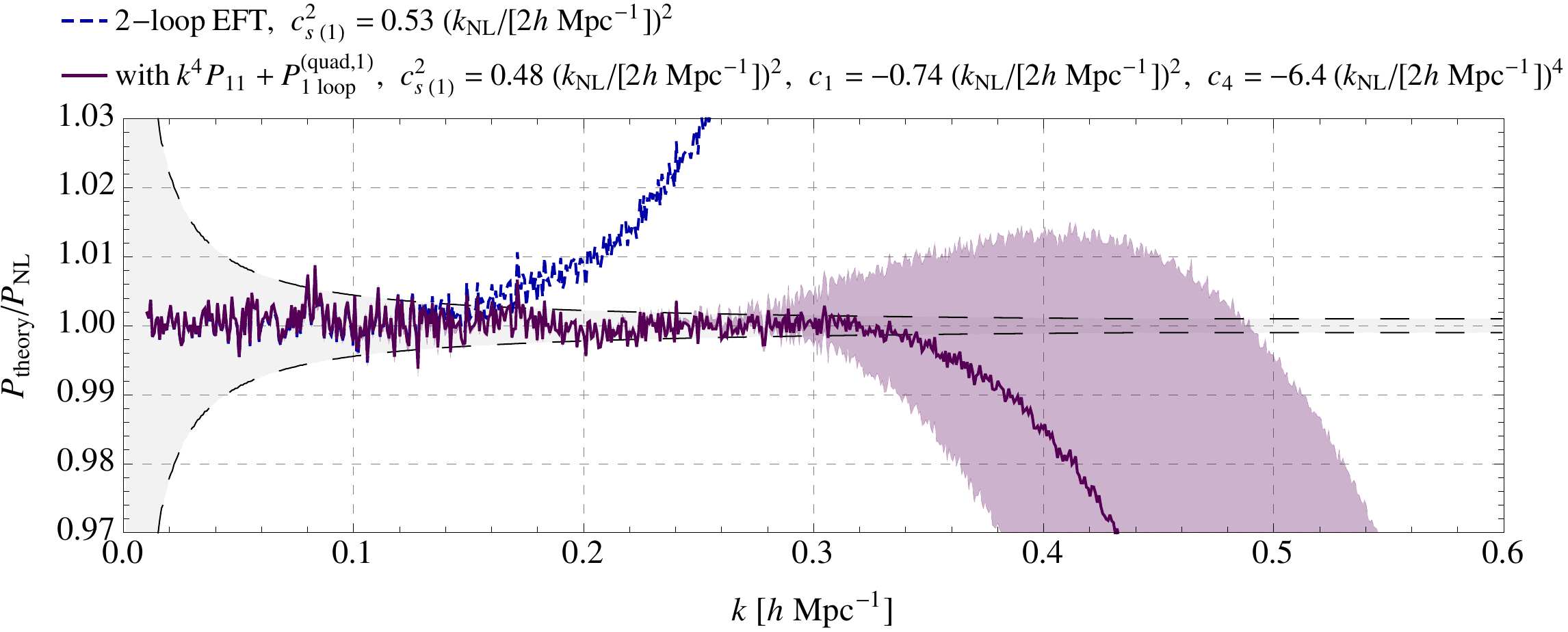}
\caption{ \small The prediction of the EFTofLSS at two loops after the inclusion of both one of the quadratic counterterms, $\sim k^2\poneloop$, and the higher-derivative counterterm, $\sim k^4P_{11}(k)$. We see that the $k$-reach is significantly improved (compared to the prediction including only one counterterm) to $\kreach\simeq 0.34\hinvMpc$, where the cosmic variance is about $10^{-3}$. The theoretical error, shaded in purple, is estimated to potentially decrease the $k$-reach to $\kreach\simeq 0.26\hinvMpc$. As discussed in Sec.~\ref{sec:UVsensitivity}, we consider this to be the most theoretically justified calculation of the EFT at two-loop order. \label{fig:both} } 
\end{center}
\end{figure}

In Fig.~\ref{fig:both}, we present a comparison of this prediction to simulation data at $z=0$, using the techniques of Sec.~\ref{sec:formulas} to fix $\ct$ and applying the fitting procedure of Sec.~\ref{sec:fittingprocedure} to the determination of $\co$, $c_1$ and $c_4$. We find that the $k$-reach is brought up to $k\simeq 0.34\hinvMpc$, where the cosmic variance is $\sim10^{-3}$, with a theoretical uncertainty that could bring the reach down to about $k\sim 0.26\hinvMpc$. The size of the cosmic variance shows the remarkable level of precision of the comparison. The parameters that we find are given by
\begin{align}
\co &\simeq 0.48\left(\knl / (2 \invMpc) \right)^2\ , \\ \nn
c_1 &\simeq -0.74\left(\knl / (2 \invMpc) \right)^2\ ,  \\ \nn
c_4 &\simeq -6.4\left(\knl / (2 \invMpc) \right)^4\ ,
\end{align}
and they are consistent with the sizes that are expected from the UV-dependence of the calculation.

\begin{figure}[t!]
\begin{center}
\includegraphics[scale=0.62]{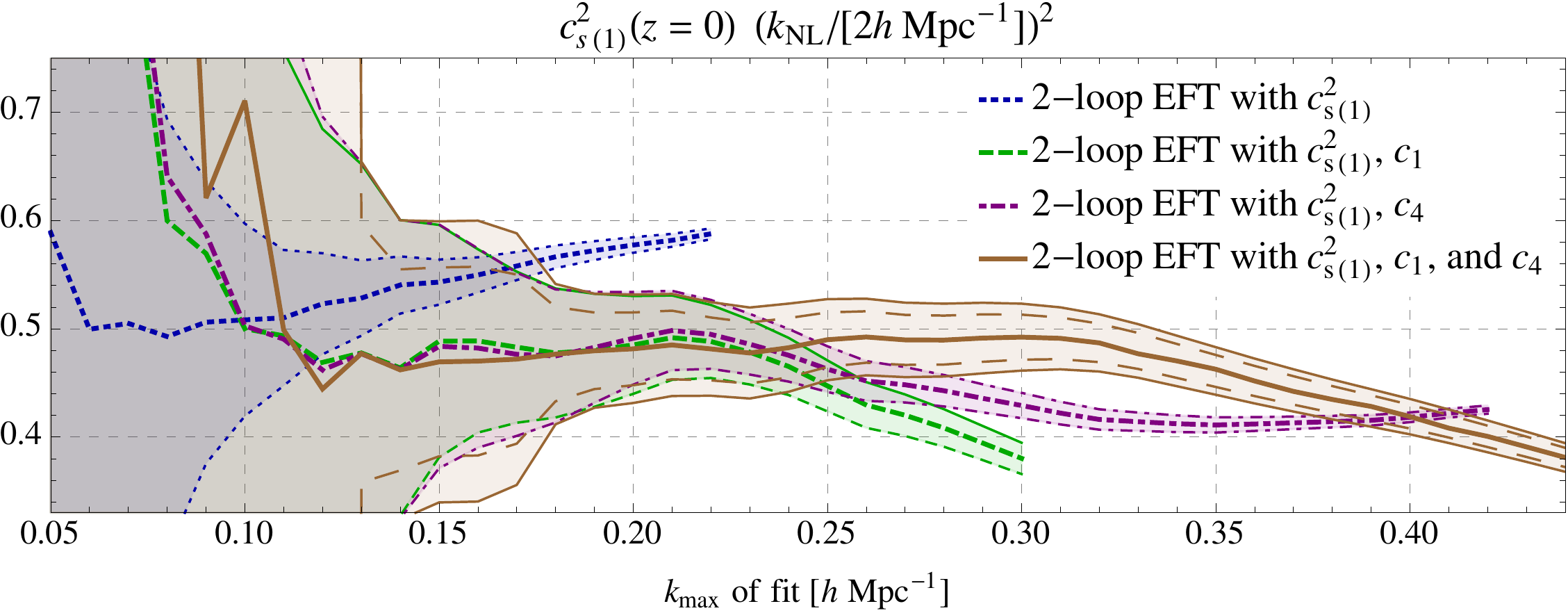}
\includegraphics[scale=0.62]{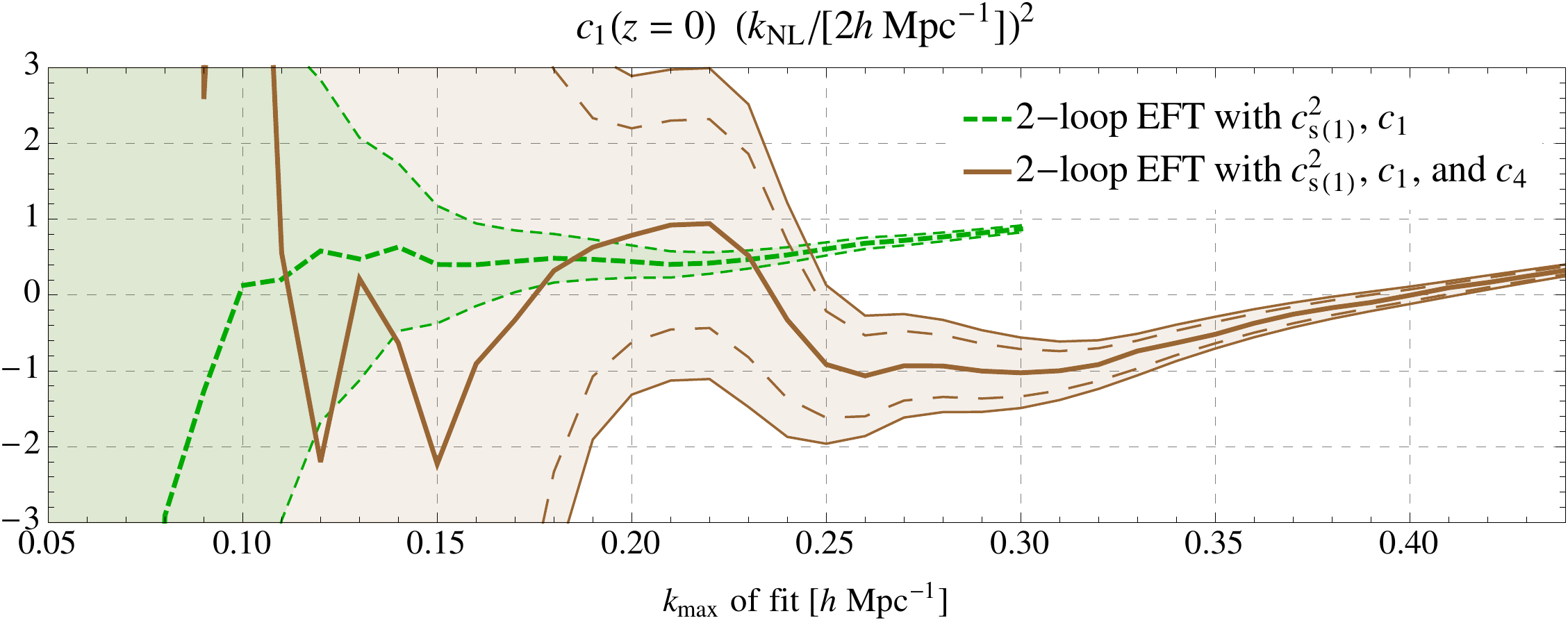}
\includegraphics[scale=0.62]{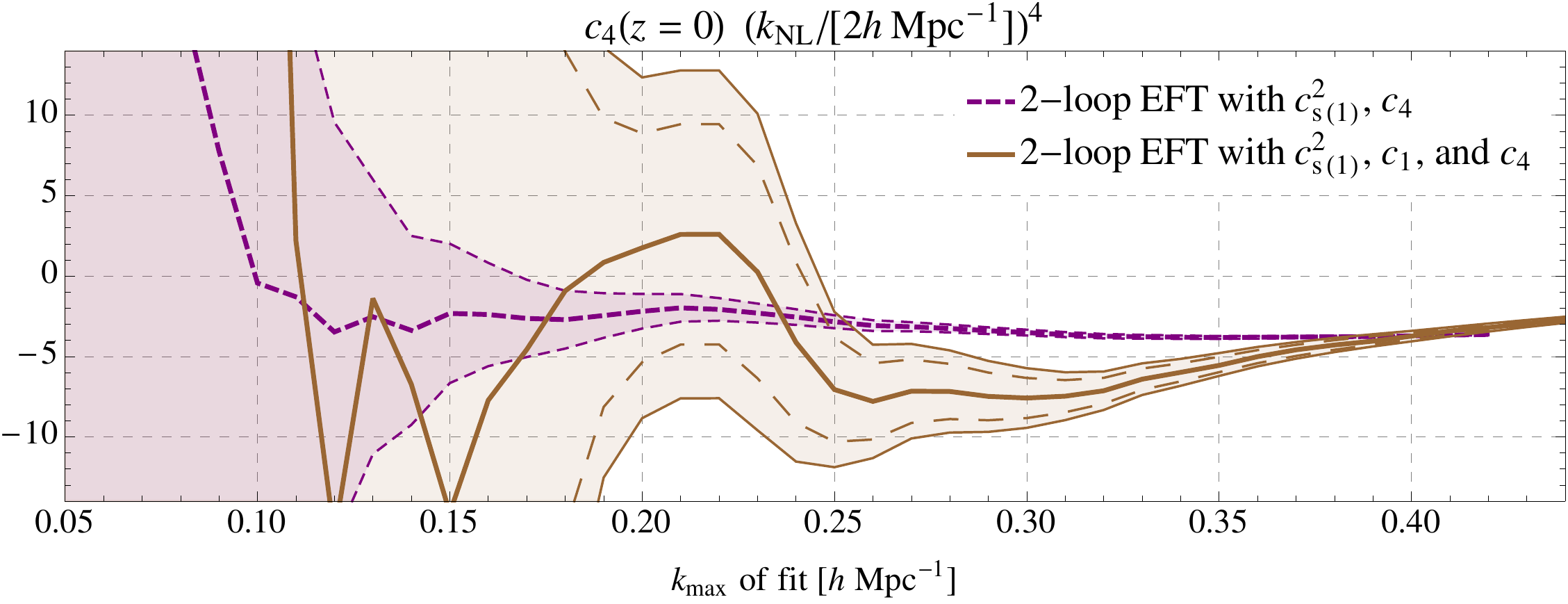}
\caption{ \small From top to bottom, values of the counterterms $\co,\;c_1$ and $c_4$ at $z=0$ as obtained from our fitting procedure as a function of the $\kmax$ of the fit. The results are presented for various choices of the counterterms being included. In shading is the $2\sigma$ errorbar from the fitting procedure, with the $1\sigma$ errorbar for the three-counterterm fit shown in long dashed lines. The presence of a flat region in $\kmax$ is interpreted as suggesting that a certain parameter is being well measured and the $\kmax$ of the fit has not been overestimated. When all counterterms are being used, we notice the presence of a flat region for all of the three parameters, ending at $\kmax\simeq 0.32\hinvMpc$. We conclude that the~$\kfit$ should be taken to be~$\simeq 0.32\hinvMpc$at $z=0$. \label{fig:flatregion} } 
\end{center}
\end{figure}

As anticipated in Sec.~\ref{sec:fittingprocedure}, an important check that we carry out in order to ensure that we are not overfitting is to ensure that the parameters that we determine from the fit are constant as we increase the fitting region. We present the results in Fig.~\ref{fig:flatregion}. In the upper figure, we can see that when we use all three counterterms, the parameter $\co$ obtained from the fit is constant over the range $k\lesssim 0.32\hinvMpc$. Here by constant we mean that the curve is constant within the error bars (shaded) determined by the fitting procedures. We also present curves for the value of $\co$ obtained when only a fraction of the necessary counterterms are included. In the blue curve, no additional counterterms are included, and the curve is never flat. 
When either $c_1$ or $c_4$ are included, the curve is flat until $k\simeq 0.22\hinvMpc $, when then it starts deviating significantly from being flat. This can be understood by noticing that at low $k$'s, the higher order counterterms are not very important, so that $\co$ can be determined very well without the complete set of the relevant counterterms being included. However, at $k\gtrsim 0.22\hinvMpc$ these terms start to become important, and it is impossible to have a flat curve for $\co$ because the term with $\co$ is trying to compensate for the lack of a relevant counterterm. 
The curves for the parameters $c_1$ and $c_4$ follow a pattern similar to the one of $\co$, being flat until $k\simeq 0.32 \hinvMpc$. The only difference is that the error bars shrink relevantly for $k \gtrsim 0.22 \hinvMpc$ because, as for the case of the $\co$ curve, these counterterms begin to be sizeable at these wavenumbers.
We conclude that if we take $\kfit=0.32\hinvMpc$, we are probably not overfitting the data. More checks on not overfitting the data are presented in Sec.~\ref{sec:checks}.

\section{Other combinations of counterterms}
\label{sec:othercounterterms}

The prediction we present in Sec.~\ref{sec:uvinsensitive} contains the minimum number of free parameters required to render the calculation UV-insensitive at the level of precision we are concerned with in this paper. Nonetheless, we can also ask what happens when we remove or add various counterterms to this prediction. This is useful for gaining insight into the effect of different combinations of terms, and for performing consistency checks. It is also useful for discovering whether there is a version of the prediction with fewer free parameters that performs just as well; in fact, even if we can {\it estimate} that some counterterms are necessary to be included to make the calculation UV-insensitive, it could well be, at least in principle, that, once we send the cutoff of the calculation to infinity, the finite contribution of the given term happens to be small and need not be included, allowing for a prediction with fewer parameters.

\subsection{Quadratic counterterms: $\sim k^2 \poneloop$}
\label{sec:quadterms}

We begin by adding the quadratic counterterms in~(\ref{eq:contribution_quad_counterterms}) to the one-counterterm two-loop EFT prediction in~(\ref{eq:peft2loop}). The functional form of each of these terms is quite similar, so we study the match of the theory to data first when one term at a time is added, then two or three terms. The results are presented in Fig.~\ref{fig:quad_terms}.

\begin{figure}[t]
\begin{center}
\includegraphics[scale=0.7]{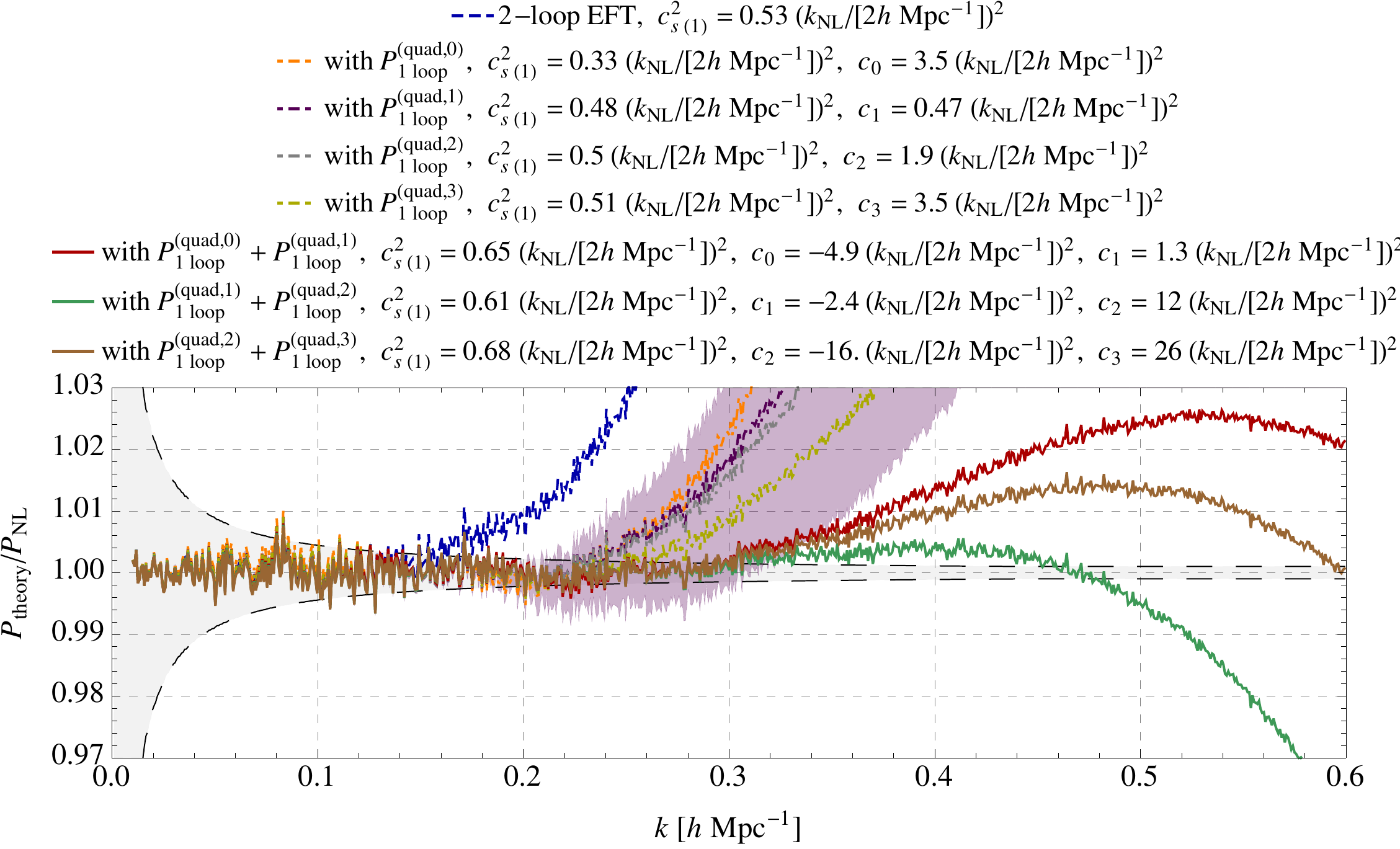}
\caption{\small Prediction of the EFT at two loops after the inclusion of the terms in~(\ref{eq:peft2loop}) plus various combinations of the quadratic counterterms. We can see that if we add just one of the counterterms, the $\kreach$ is increased to $\kreach\simeq 0.23-0.26\hinvMpc$, depending on the counterterm that is chosen, where the cosmic variance is about $\sim2\times 10^{-3}$. An estimate of the theoretical error is shaded in purple, showing the possibility of the decrease of the $\kreach$ all the way to $\kreach\simeq 0.18\hinvMpc$.  When we include two of the quadratic counterterms, the reach of the theory can be further increased to $\kreach=0.3\hinvMpc$, although this is likely the result of overfitting, as described in the main text.
}\label{fig:quad_terms}
\end{center}
\end{figure}
 
When we add just one of the counterterms, we see that the EFTofLSS matches the data up to $k_{\rm reach} \simeq 0.23-0.26\hinvMpc$, depending on the counterterm we use, where the cosmic variance of the data is as small as $2\times 10^{-3} $. The sizes of the numerical prefactors when each term is included separately are given respectively by
\bea
&&c_0\simeq 3.5\left(\knl / (2 \invMpc) \right)^2\ ,\\ \nn
&&c_1\simeq 0.47\left(\knl / (2 \invMpc) \right)^2\ ,\\ \nn
&& c_2\simeq 1.9\left(\knl / (2 \invMpc) \right)^2\ ,\\ \nn
&& c_3\simeq 3.5\left(\knl / (2 \invMpc) \right)^2\ ,
\eea
while $\co$ is in the range $0.33\left(\knl / (2 \invMpc) \right)^2 \lesssim \co \lesssim 0.51\left(\knl / (2 \invMpc) \right)^2$, with slight variations as we add more terms. As expected from the analysis of the bispectrum~\cite{Angulo:2014tfa}, we see that for these terms to be relevant, their coefficients need to be boosted with respect to the value of $\co$ by large factors. The value of $c_1$ is therefore compatible with what is expected from the UV sensitivity of $\ptwoloop$ in~(\ref{eq:UV-coeff})~\footnote{As we stressed, the estimate of the induced size of a counterterm from UV physics should not be overinterpreted as more than what it is, a rough indication of the expected numerical value. In particular, we argued that  $c_4$ should also be included in the calculation, but its inclusion is expected just to provide an order one correction to the size of $c_1$ obtained from the fit to the data.}. 
 
 When we add two of these quadratic counterterms at the same time, we see that the reach of the theory can be boosted to as much as $k_{\rm reach}=0.3\hinvMpc$. However, we interpret this higher match of the theory with data as an overfit of the theory. In fact, in some cases, this higher reach is achieved by boosting the coefficients of the counterterms to what we interpret to be very large values, in such a way that the two contributions cancel each other, when on the contrary these contributions are not expected to be canceling against each other. This interpretation is confirmed by the fact the numerical coefficients that we obtain when we add two counterterms become quite large with respect to $\co$ (this is particularly true for $c_2$ and $c_3$)  and to what is roughly expected from the UV in (\ref{eq:UV-coeff}); for example, when $P^\text{(quad,\,1)}_\text{1-loop}$ and $P^\text{(quad,\,2)}_\text{1-loop}$} are both included, we find that
   \begin{align}
    \co &\simeq 0.61\left(\knl / (2 \invMpc) \right)^2\ , \\ \nn
 c_1 &\simeq -2.4\left(\knl / (2 \invMpc) \right)^2\ ,\\ \nn
 c_2 &\simeq 12\left(\knl / (2 \invMpc) \right)^2\ ,
 \end{align}
and most importantly they change a lot with respect to their numerical values when only one term was used. Furthermore, they take opposite signs so that they can indeed cancel each other. In the case of $c_0$ and $c_1$, it is mainly the mismatch from what is expected from the UV that pushes towards the interpretation of the increase of the $\kreach$ as an overfit.  These considerations tell us that to ensure that we are not overfitting the data we should not rely only on the consistency of the measurement of the parameters as a function of $\kfit$, but also on the estimated size of a term from the UV and on the change of the value of the terms as we include additional parameters.
 
 When we add three counterterms, and we impose that the value of $\co$ is not changed by more than a factor of two with respect to the value that we find  when not including these terms (a fact that is justified by the hierarchy of the various contributions), we find that the $k$-reach of the theory is not relevantly improved, so we have not plotted these curves in Fig.~\ref{fig:quad_terms}.  We therefore conclude that for the precision of the given data, and restricting only to the quadratic counterterms, it is enough to include only one of them, where the reach of the theory is boosted to about $k_{\rm reach} \simeq 0.23\hinvMpc$. This $\kreach$ is inferior to what is obtained in the consistent calculation that was presented earlier, where we include also the higher derivative counterterm. 
 
 Finally, we make an additional comment. As noticed in~\cite{Carrasco:2013mua,Carroll:2013oxa}, the EFTofLSS is local in space, but non-local in time. This means that the term $\poneloopcs$ should actually correspond to three different terms, with slightly different functional forms (see~\cite{Foreman:2015uva} for the most clear presentation). We have checked that the functional form of each of these terms is however highly degenerate with the one from the quadratic counterterms and the $k^2P_{11}(k)$ term; and in fact, when we use them in the fit to replace the quadratic counterterms, the result is not significantly different. Similarly, in the non-local-in-time treatment, each of the quadratic counterterms leads to two independent functional forms, but we have checked that they are also degenerate with the quadratic counterterms in the local-in-time approximation and the $k^2P_{11}(k)$ term.

\subsection{Four-derivative  counterterm: $\sim k^4\,P_{11}$}

\begin{figure}[t]
\begin{center}
\includegraphics[scale=0.7]{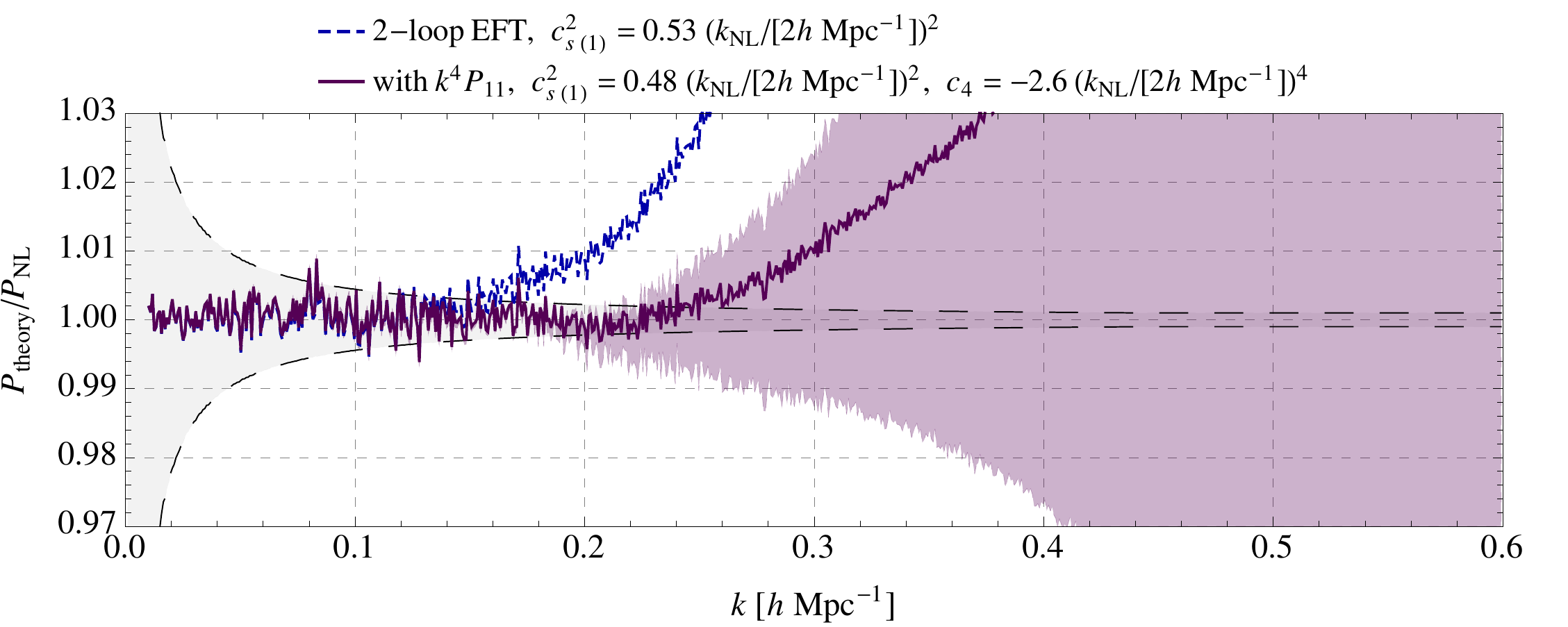}
\caption{ \small The prediction of the EFTofLSS at two loops after the inclusion of the higher-derivative counterterm $k^4P_{11}(k)$. We see that the $\kreach$ is improved to $\kreach\simeq 0.24\hinvMpc$, quite similar to the increase we saw in Sec.~\ref{sec:quadterms} when adding a single quadratic counterterm.   }\label{fig:higher-deriv}
\end{center}
\end{figure}

We next pass to study the effect of including only the higher-derivative counterterm~(\ref{eq:highr_contribution}) in the two-loop calculation with the $\co$ counterterm only.
The result is presented in Fig.~\ref{fig:higher-deriv}. We see that when we add this term alone the $k$-reach of the EFT is boosted to $k\simeq 0.24\hinvMpc$. The parameter values that we find are 
\be
\co\simeq 0.48\left(\knl / (2 \invMpc) \right)^2\ ,\quad  
c_4\simeq -2.6\left(\knl / (2 \invMpc) \right)^4\  ,\quad 
\ee
which are compatible with what is expected from (\ref{eq:UV-coeff}). We see that the inclusion of this term results in a $\kreach$ that is very similar to that obtained from adding a single quadratic counterterm in Sec.~\ref{sec:quadterms}. Evidently, either a single quadratic term or a single higher-derivative term are not sufficient to reproduce the results of adding both terms in Fig.~\ref{fig:both}.

\subsection{Stochastic counterterm: $\sim k^4$}

\begin{figure}[t]
\begin{center}
\includegraphics[scale=0.7]{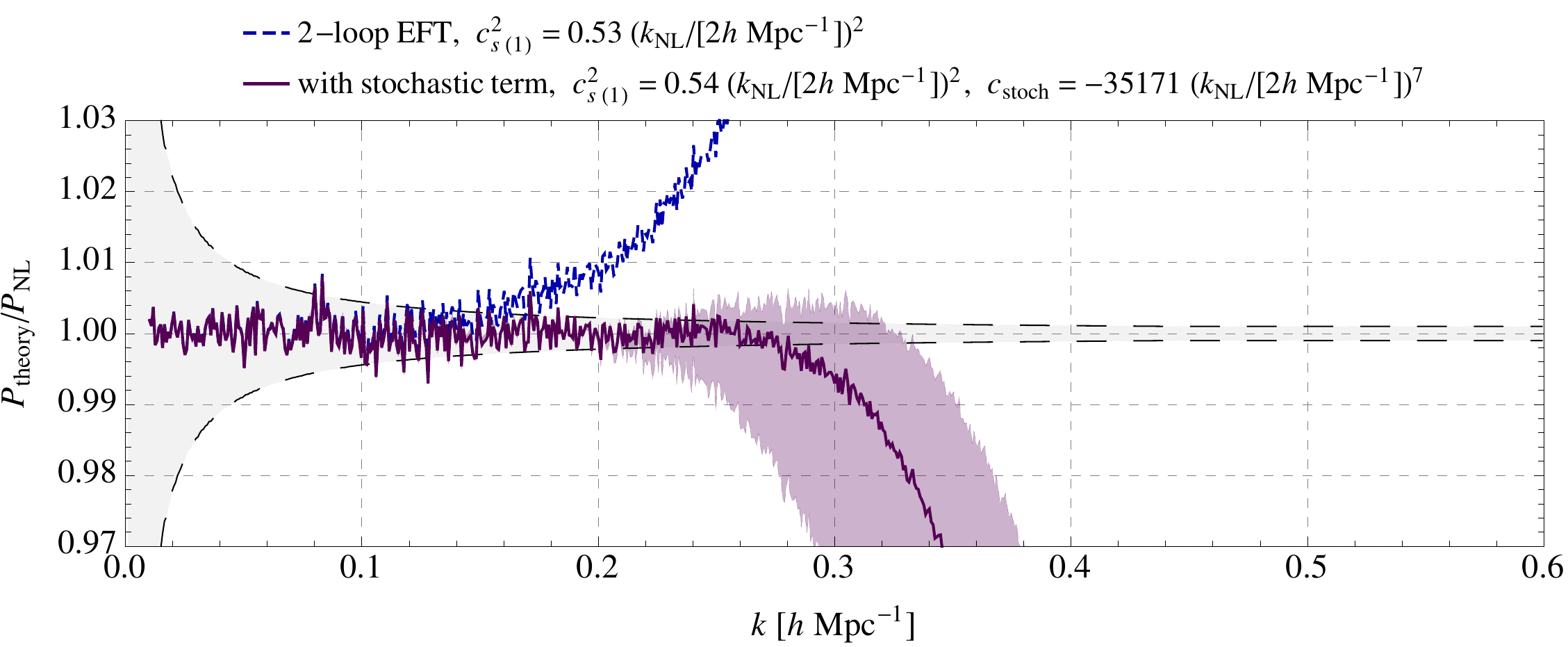}
\caption{\small The prediction of the EFTofLSS at two loops with the addition of the stochastic counterterm. We find that the $\kreach$ is improved to about $\kreach\simeq0.34\hinvMpc$, with a theory error, in purple, that could decrease the reach to $\kreach\sim 0.21\hinvMpc$. However,  the magnitude and the sign of the required stochastic counterterm seem  to be in conflict with the theoretical expectations, as explained in the main text.  }\label{fig:stoch}
\end{center}
\end{figure}

Finally, we consider the addition of a stochastic counterterm~(\ref{eq:P_stoch}) to the one-parameter two-loop formula from~(\ref{eq:peft2loop}).
Naively, we expect this term to contribute only at a much higher order in perturbation theory than the order at which we are working. However, there is subtlety in determining the expected size of this counterterm, as first pointed out in~\cite{Angulo:2015eqa} in the context of biased tracers. An equivalent way of writing (\ref{eq:P_stoch}) is
\be\label{eq:P_stoch2}
P_\text{stoch}(k)\sim (2\pi)^2\left(\frac{k}{k_{\rm M}}\right)^4\frac{1}{\bar n} \ .
\ee
Here $\bar n$ is the number density of the objects that most contribute to the stochastic noise. The derivatives acting on this term are expected to be suppressed by the inverse length scale of the same objects, which we call $k_{\rm M}$. Now, for the dark matter power spectrum we expect both of these scales to be of order $\knl$, which is why (\ref{eq:P_stoch}) is written with $\cstoch$ expected to be an order one number. In particular, this is true for the part of the counterterm that is supposed to correct the perturbative loops, which have no information about the non-perturbative halos. However, given the high number of powers in which these scales appear, one should be careful, because the small difference in these scales from $\knl$ can make quite a difference. For example, if very massive halos are the ones that contributes, $1/\bar n$ is very large and $k_{\rm M}$ is very small; vice-versa if it is small halos that contribute. Therefore, the value of~$\cstoch$ from~(\ref{eq:P_stoch}) could indeed be very far from order one, and this would affect at which order in perturbation theory the term becomes relevant.

The result of adding a stochastic counterterm to the prediction from~(\ref{eq:peft2loop}) is shown in Fig.~\ref{fig:stoch}. We can see that adding the stochastic counterterm makes the EFT match the data up to $k\simeq 0.28\hinvMpc$, where the cosmic variance of the simulation is about $3\times 10^{-3}$. This is worse than the full UV-insensitive prediction, and not much better than even adding a single quadratic counterterm. The numerical values of the parameters from this fit are
\be
\label{eq:stoch-parameter}
\co=0.54\left(\knl / (2 \invMpc) \right)^2 \ , \quad 
\cstoch=-3.5\times 10^4 \left(\knl / (2 \invMpc) \right)^7\ .
\ee

Although the magnitude of $\cstoch$ is comparable to what is required to fix the  UV contribution to $\ptwoloop$, it is not compatible because it has the wrong sign. Indeed, the fitted value of $\cstoch$, negative in our case, is composed of a sum of UV and \emph{finite} contributions. As seen in \eqref{eq:UV-coeff}, the UV contribution requires a positive $\cstoch$. Furthermore, as we will explain now, the finite contribution to $\cstoch$ has to be positive as well, so that the overall sign of $\cstoch$ is expected to be positive.

\begin{figure}[t]
\begin{center}
\includegraphics[scale=0.5]{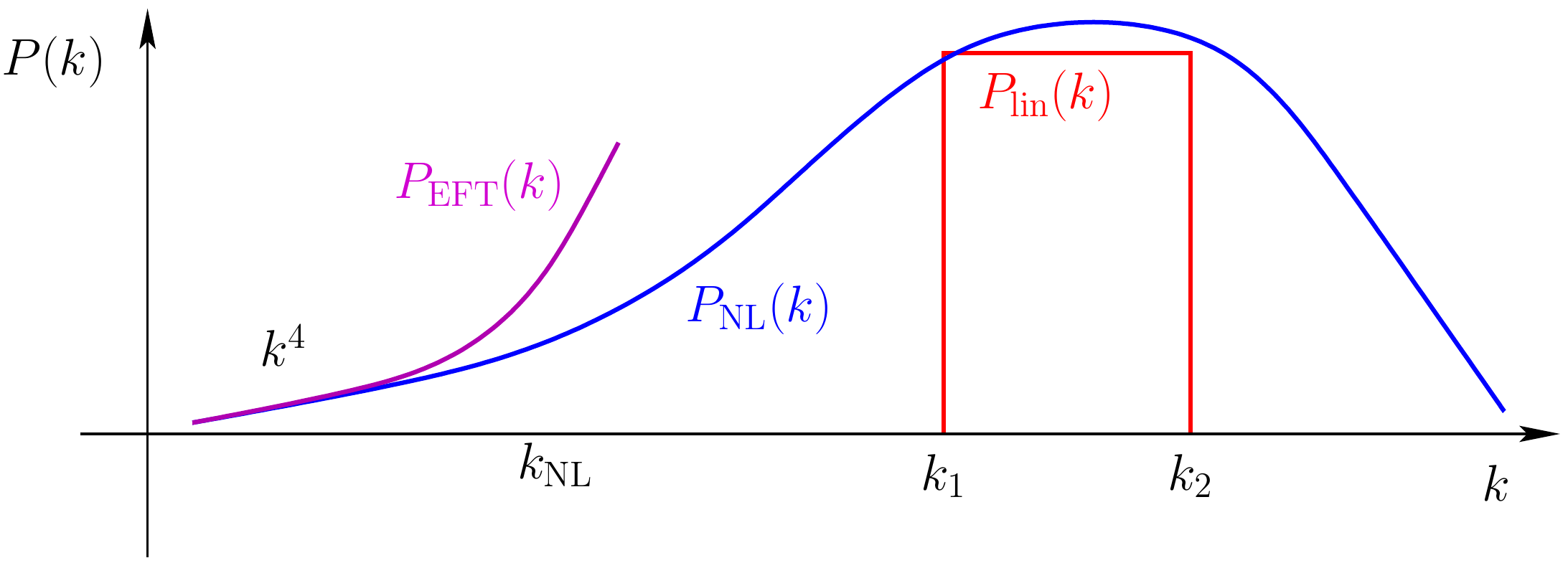}
\caption{\small Sketch of the nonlinear power spectrum in a toy model where the linear power spectrum has only short scale power. The prediction of the EFT as $k\to 0$ is $P_{\rm EFT}^{\rm toy} \sim (2\pi)^2 \cstoch\left(k/\knl\right)^4 \knl^{-3}$. Since the power spectrum must be positive for all $k$, this implies that $\cstoch>0$. }\label{fig:toy}
\end{center}
\end{figure}

Let us imagine a toy universe (shown schematically in Fig.~\ref{fig:toy}) where the linear power spectrum is non-vanishing only over a small region between $k_1\leq k\leq k_2$. At a time late enough so that $\knl<k_1$, the vanishing of $P_{11}(k)$ in the perturbative regime implies that the EFT prediction for the power spectrum is very simple:
\be
P_{\rm EFT}^{\rm toy} \sim (2\pi)^2 \cstoch\left(\frac{k}{\knl}\right)^4 \frac{1}{\knl^3} 
+(2\pi)^2\cstoch^{(2)}\left(\frac{k}{\knl}\right)^6 \frac{1}{\knl^3} + \ldots \ .
\ee
Remarkably, in this toy universe the prediction at long wavelengths is entirely dominated by the stochastic contributions! Since the power spectrum is the expectation value of $|\delta_k|^2$, it must be positive for all $k$. By taking the limit $k\to 0$, we conclude that $\cstoch\geq 0$. Notice that this argument requires that $\cstoch$ is the {\it finite} contribution. The moment $\cstoch$ includes a renormalization of a loop, we cannot make this argument any longer. But, as we discussed, this is not the case at hand.

We therefore conclude that we have no evidence of the necessity of adding a stochastic counterterm before the other counterterms in the UV-insensitive calculation~\footnote{It is also reasonable to ask whether the coefficient of the stochastic term is so large that it should be included before any two-loop term, but after the one-loop terms from~(\ref{eq:peft1loop}). When we fit such a formula to the data, however, we find that the $k$-reach is not significantly improved over the one-parameter one-loop prediction, and furthermore, the sign of $c_\text{stoch}$, which again is just given by the finite contribution, is negative, just as for the ``two-loop+stochastic" prediction.  }. We have tried to add the stochastic term after the inclusion of just the quadratic counterterm associated with $c_1$, finding the same conclusions as when we add the stochastic term without the $c_1$ term. It is possible that the stochastic term should be added after both the $c_1$ and $c_4$ terms are incuded, where the EFTofLSS stops fitting the nonlinear power spectrum due to lack of power. It is also possible that the stochastic counterterm might play  a relevant role before the $\kreach$ of the calculation with $c_1$ and $c_4$, so that, without the stochastic term, the additional terms would lead to a lack of power in the EFT prediction at a lower wavenumber. Though this is possible, we have no evidence of this from the fit to the power spectrum until the estimated $\kreach$ of the computation. But in either of these cases, it is likely that one should evaluate the three-loop contribution first.

\section{Higher Redshifts}

We now proceed to study redshifts higher than $z=0$. This will be useful to explore how much the $\kreach$ of the EFTofLSS is improved as we move to higher redshifts, and also to explore the time dependence of the counterterms.

\subsection{Fits to the power spectrum }

\begin{figure}[t!]
\begin{center}
\includegraphics[scale=0.76]{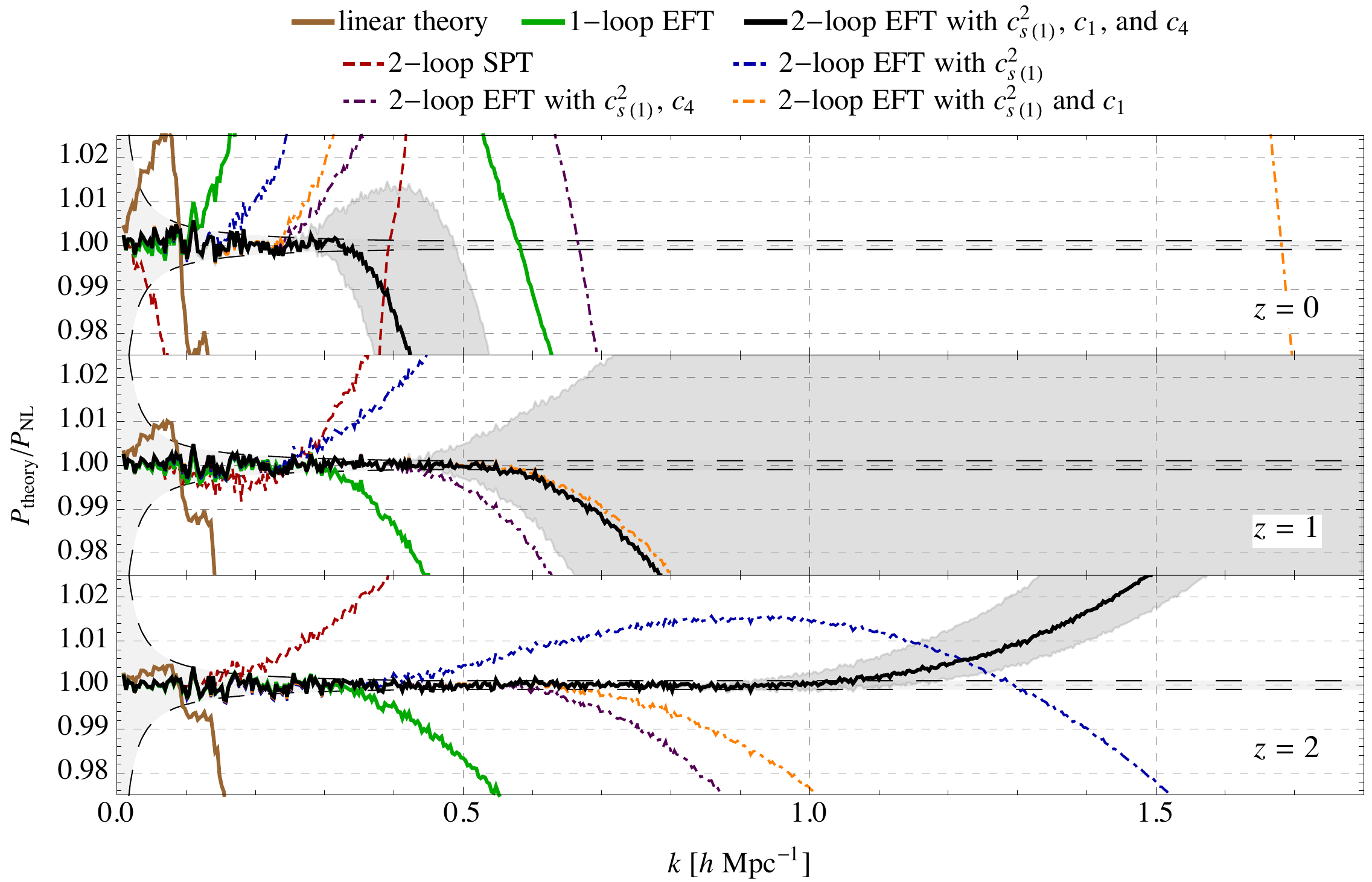}
\caption{\small The prediction of the EFTofLSS for the matter power spectrum as a function of redshift. In black, we plot the EFTofLSS with three counterterms, with parameters $\co$, $c_1$, and $c_4$ fit separately at each redshift. The darker grey band corresponds to an estimate of the theoretical error estimated by taking the value of $\co$ which is $1\sigma$ off from the best fit obtained at $0.75\kfit$. We also plot two-loop EFTofLSS predictions with different combinations of counterterms, and various other lower-order predictions. We see that the $\kreach$ is higher and higher with the higher redshifts, and the gain with respect to SPT is very substantial at all redshifts. In is expected that the $\kreach$ as a function of redshift is a smooth function of $z$, once we take the theoretical error in account.}\label{fig:allz}
\end{center}\end{figure}

The results of applying the same procedure that we described at $z=0$ to higher redshift are given in Fig.~\ref{fig:allz}. Figures of the values of the counterterms as a function of $\kmax$, from which we determine $\kfit$ and the theoretical error, are provided in App.~\ref{app:fitvalue}. When we consider the calculation done with the three relevant counterterms (with parameters $\co$, $c_1$, and $c_4$), we clearly see that as we move to higher redshifts, the $k$-reach is relevantly improved, to  $k\simeq 0.6\hinvMpc$ at $z=1$ and $k\simeq 1.1\hinvMpc$ at $z=2$.   Based on our estimates of the theoretical error in the fits, the $k$-reach of the prediction could potentially be as low as $k\simeq0.26\hinvMpc$ at $z=0$,  $k\simeq0.4 \hinvMpc$ at $z=1$, and as low as $k\simeq0.9 \hinvMpc$ at $z=2$. Nevertheless, the EFT provides a substantial gain with respect to other analytical techniques, such as SPT at two loops (also shown in Fig.~\ref{fig:allz}). Such a gain becomes very important once we consider that the number of available modes scales as $\kreach^3$. For example, at redshift $z=1$, the gain in number of modes with respect to two-loop SPT is about 200. At $z=2$, this same number is closer to $400$~\footnote{In particular, one can notice that SPT fails to match the data at such low wavenumbers that cosmic variance plays a relevant role in determining its $\kreach$. A more accurate estimate of its $\kreach$ can be obtained by noticing when SPT significantly deviates from the EFT predictions.}.

\subsection{Additional checks of fitting procedure}
\label{sec:checks}

Given that we are fitting the power spectrum, albeit with very small error bars, with three parameters, there is a certain concern that we might be overfitting, despite the fact that we have designed the fitting procedure described in Sec.~\ref{sec:fittingprocedure} to minimize this possibility. We try to limit the possibility of overfitting by performing the following additional checks. First, as mentioned, we present a theoretical error on the prediction. Second, we have verified that the functional forms of the various counterterms do not cancel each other relevantly~\footnote{In~\cite{Foreman:2015uva} instead it was noticed that by pushing $\kren$ to higher values, there was some cancellation among two-loop diagrams and the $\poneloopcs$ counterterm. This cancellation would have disappeared if some slightly different value of~$\co$ was chosen, implying a lower $\kreach$ of the theory. See Sec.~\ref{sec:z0comp} for further discussion.}. Third, we checked that if we try to fit the numerical data by setting $\ptwoloop=0$, we are unable to  match the numerical data as successfully as when we include $\ptwoloop$. 

 Finally, we check that the numerical values of the parameters we obtain are consistent with the size that is induced by the uncontrolled UV. This can be estimated by repeating the procedure used to obtain (\ref{eq:UV-coeff}), with the only difference that the counterterms are estimated by fitting the difference of $\ptwoloop$ computed with cutoff $\Lambda=\infty$ and a $z$-dependent $\Lambda(z)$ that roughly approximates what the nonlinear scale is expected to be as a function of redshift.  In more detail, we choose the cutoff to be $\Lambda(z)=2\,\kreach(z)$ and fit the counterterms to 
 $P_\text{2-loop}^{\Lambda=\infty}(k,z)-P_\text{2-loop}^{\Lambda=2\,\kreach(z)}(k,z)$
 over the $k$-range $0.05 \hinvMpc - 0.5\kreach(z)$ including a $0.3\%$ error bar on the computation of the integrals. We obtain the following values for the counterterms for $z=\{0,1,2\}$ when including the counterterms associated to $c_1$, $c_4$ and $\cstoch$\footnote{
If instead we choose $\Lambda = 1\, k_{\rm reach}(z)$ we obtain
\bea
\label{eq:UV-coeffz}
&& c_{1}^\text{(UV)}(z=\{0,1,2\}) = \{-2.8,-0.73,-0.18 \}  \left(\knl /( 2 \invMpc) \right)^2 \ ,\\ \nonumber 
&& c_{4}^\text{(UV)}(z=\{0,1,2\}) = \{-13,-1.2,-0.14\}  \left(\knl /( 2 \invMpc) \right)^4 \ ,\\ \nonumber 
&& \cstoch^\text{(UV)}(z=\{0,1,2\}) = \{1.4 \times 10^{5},5.6 \times 10^3,340\}  \left(\knl /( 2 \invMpc) \right)^7 \ ,
\eea
and 
\bea
&& c_{1}^\text{(UV)}(z=\{0,1,2\}) = \{-3.9,-1.1,-0.20 \}  \left(\knl /( 2 \invMpc) \right)^2 \ ,\\ \nonumber 
&& c_{4}^\text{(UV)}(z=\{0,1,2\}) = \{-17,-1.7,-0.15\}  \left(\knl /( 2 \invMpc) \right)^4 \ , \nonumber 
\eea
when fitting only the terms associated to $c_1$ and $c_4$.
For $\poneloop(k,z)$ we have
\begin{equation}
c_{s(1)}^\text{2,(UV)}(z=\{0,1,2\}) = \{-1.2,-0.11,-0.04 \}  \left(\knl /( 2 \invMpc) \right)^2.
\end{equation}

 } 
\bea
\label{eq:UV-coeffz}
&& c_{1}^\text{(UV)}(z=\{0,1,2\}) = \{-1.1,-0.27,-0.064 \}  \left(\knl /( 2 \invMpc) \right)^2 \ ,\\ \nonumber 
&& c_{4}^\text{(UV)}(z=\{0,1,2\}) = \{-5.3,-0.46,-0.050\}  \left(\knl /( 2 \invMpc) \right)^4 \ ,\\ \nonumber 
&& \cstoch^\text{(UV)}(z=\{0,1,2\}) = \{6.2 \times 10^{4},2.4 \times 10^{3},130\}  \left(\knl /( 2 \invMpc) \right)^7 \ ,
\eea
and 
\bea
&& c_{1}^\text{(UV)}(z=\{0,1,2\}) = \{-1.6,-0.47,-0.092 \}  \left(\knl /( 2 \invMpc) \right)^2 \ ,\\ \nonumber 
&& c_{4}^\text{(UV)}(z=\{0,1,2\}) = \{-7.0,-0.74,-0.068\}  \left(\knl /( 2 \invMpc) \right)^4 \ , \nonumber 
\eea
when fitting only the terms associated to $c_1$ and $c_4$.
Applying the same method to $\poneloop(k,z)$ (fitting from $k_{\rm min}=0.005 \hinvMpc$) we have
\begin{equation}
c_{s(1)}^\text{2,(UV)}(z=\{0,1,2\}) = \{-2.2,-0.28,-0.11 \}  \left(\knl /( 2 \invMpc) \right)^2.
\end{equation} 

These values are in general not very different from the coefficients that we find fitting the power spectrum numerical data in Table~\ref{tab1} and the values of $c_1$ and $c_4$ are quite independent of the presence of the stochastic term.  Note that the values of the parameters are also quite independent from the choice of $\Lambda$, not varying more that a factor of 2-3 when considering the range $\Lambda = 1-2 \,k_{\rm reach}(z)$. They are also stable under the change of the fitting range as one varies $k_{\rm max} = 0.2-0.9\, k_{\rm reach}$. One also  notices that while at $z=1$ the quadratic counterterm seems to bring most of the $k$-gain, this is not the case at $z=0$ and $z=2$. Thus, along with the arguments about UV-sensitivity, the data themselves seem to indicate that the inclusion of all three counterterms is the most appropriate choice. 
   
   In summary, we find these checks to be quite successful. We conclude that we find no strong indications that we are overfitting, even though we acknowledge that a better determination of the value of the counterterms by analyzing higher statistics  or observables, as done for example in~\cite{Senatore:2014via,Angulo:2014tfa,Baldauf:2015tla}~\footnote{We notice that these additional ways to determine  the value of the counterterms are possible because the counterterms of the EFTofLSS are terms that appear in some equations of motion and for which we know their origin in terms of UV degrees of freedom. This implies the fact that the same parameter appears in multiple observables or that one can measure them using directly dark matter particles. This is one of the characteristics because of which the EFTofLSS, being a theory and not a model, is more predictive than other approaches.}, or with direct measurement from small $N$-body simulations~\cite{Carrasco:2012cv}, or by including higher order terms, would be helpful.

   \begin{table}[t]
\begin{center}
   \begin{tabular}{| c || c | c | c |   | c | c | c |    | c | c | c ||}
   \hline
& \multicolumn{3}{|c||}{$z=0$} & \multicolumn{3}{|c||}{$z=1$} & \multicolumn{3}{|c||}{$z=2$} \\
\hline
    & $\co$ & $c_1$ & $c_4$ &  $\co$ & $c_1$ & $c_4$ &     $\co$ & $c_1$ & $c_4$ \\
   \hline
only $\co$ &  0.53 &  x &  x &  0.20  &  x &  x  & 0.073  &  x &  x\\
   \hline
only $\co$ \& $c_1$ & 0.48 &  0.47 & x & 0.18  &  0.23  & x  & 0.066 & 0.063  & x\\
   \hline
only $\co$ \& $c_4$ & 0.48  &  x & -2.55 &  0.19  &  x & -0.29 &  0.069 &  x &  -0.033\\
   \hline
 $\co$ \& $c_1$ \& $c_4$ & 0.48 & -0.74 & -6.41 & 0.18 & 0.22 & -0.014 & 0.060 & 0.15 & 0.040 \\
   \hline 
\end{tabular}
\caption{\small  Table of the numerical values of the counterterms as a function of redshift $z$ and for the various combinations that are studied in the paper. The units are $(\knl/(2 \hinvMpc ))^2$ for $\co$ and $c_1$ and $(\knl/(2 \hinvMpc))^4$ for $c_4$.}\label{tab1}
   \end{center}
   \end{table}

   \subsection{Time-dependence of counterterms}

Since time translations are spontaneously broken in the universe, the time-dependence of the EFT parameters is unknown. The timescale of these coefficients is expected to be of order Hubble, so there should exist approximate functional forms related to this timescale. For each counterterm, we present a quasi-two-parameter fitting function that works reasonably well.

\begin{figure}[t]
\begin{center}
\includegraphics[scale=0.9]{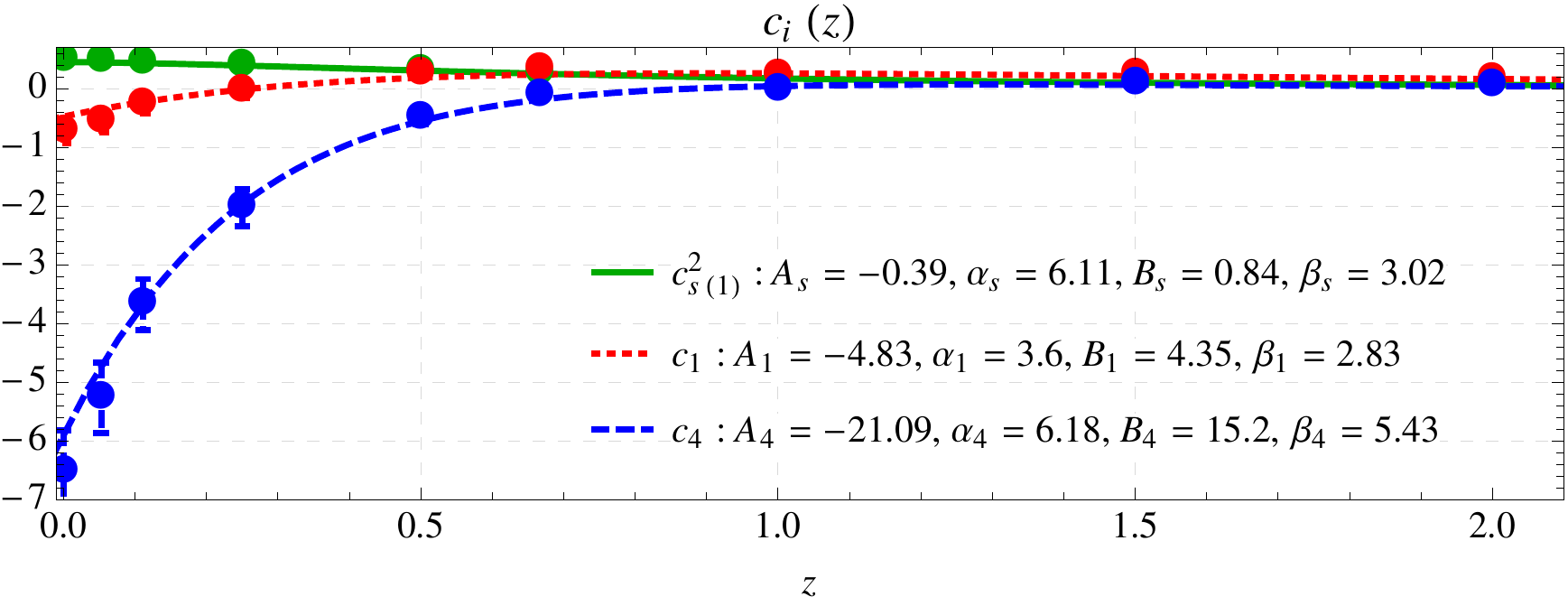}
\caption{\small 
 {Time dependence of $\co(z)$, $c_1$ and $c_4¨$ according to~(\ref{eq:three-param}), where $A_i$ and $\alpha_i$ are constrained to be close to the values obtained from fitting (\ref{eq:UV-coeffz}).}
}
\label{fig:csz}
\end{center}\end{figure}

We parametrize the time-dependence of each counterterm as the sum of two power laws in the following way
\begin{align}
\label{eq:three-param}
\co(z) &= 	A_{s}\, D_1(z)^{\alpha_{s}}+B_{s}\, D_1(z)^{\beta_{s}}\ , \\ \nn 
c_1(z) &=	A_{1}\, D_1(z)^{\alpha_{1}}+B_{1}\, D_1(z)^{\beta_{1}}\ , \\ \nn 
c_4(z) &=	 A_{4}\, D_1(z)^{\alpha_{4}}+B_{4}\, D_1(z)^{\beta_{4}}\ .
\end{align}
The first power law, characterized by $A_i$ and $\alpha_i$, represents the expected time-dependence of the counterterm induced from the cancellation of UV part of the loops, while the second power law, characterized by $B_i$ and $\beta_i$, is expected to be associated to the finite terms. For each of the three counterterms $\co$, $c_1$, and $c_4$, we fit these four coefficients. However, we call this a quasi-two parameter fit because we constrain the values of  $A_i$ and $\alpha_i$ to lie close to the values obtained from fitting the time-dependence of the UV coefficients from (\ref{eq:UV-coeffz})  by a power law. The time-dependence of the parameters depends on which counterterms are included in the power spectrum fits, as this changes the reach of the theory, which in turn changes the cutoff $\Lambda(z)$ used in the determination of $\co$, $c_1$, and $ c_4$. Here we use the $k$-reach obtained by including both the counterterms associated to $c_1$ and $c_4$ as we believe that they are the ones that need to be included. We fix the range that $A_i$ can take to be between $0.33$ and $3.0$ times the best fit of the UV part of the parameters~$c_i$, and $\alpha_i$ to lie within $0.75$ and $1.33$ times the best fit. These values were chosen by analyzing the change in the best fit values occurring when varying the ratios $\Lambda(z) / \kreach(z)$ from $1$ to $3$ and also from the change in their values when changing the fitting range from $k=0.05\invMpc$ to $0.2 \, \kreach(z)$ to  $k=0.05 \hinvMpc$ to $0.9 \Lambda(z)$. We believe that these represent reasonably well the uncertainty on the determination of the UV contribution to $\co$, $c_1$, and $c_4$ .
The results of the fits are presented in Fig.~\ref{fig:csz}.

\begin{figure}[t]
\begin{center}
\includegraphics[scale=0.76]{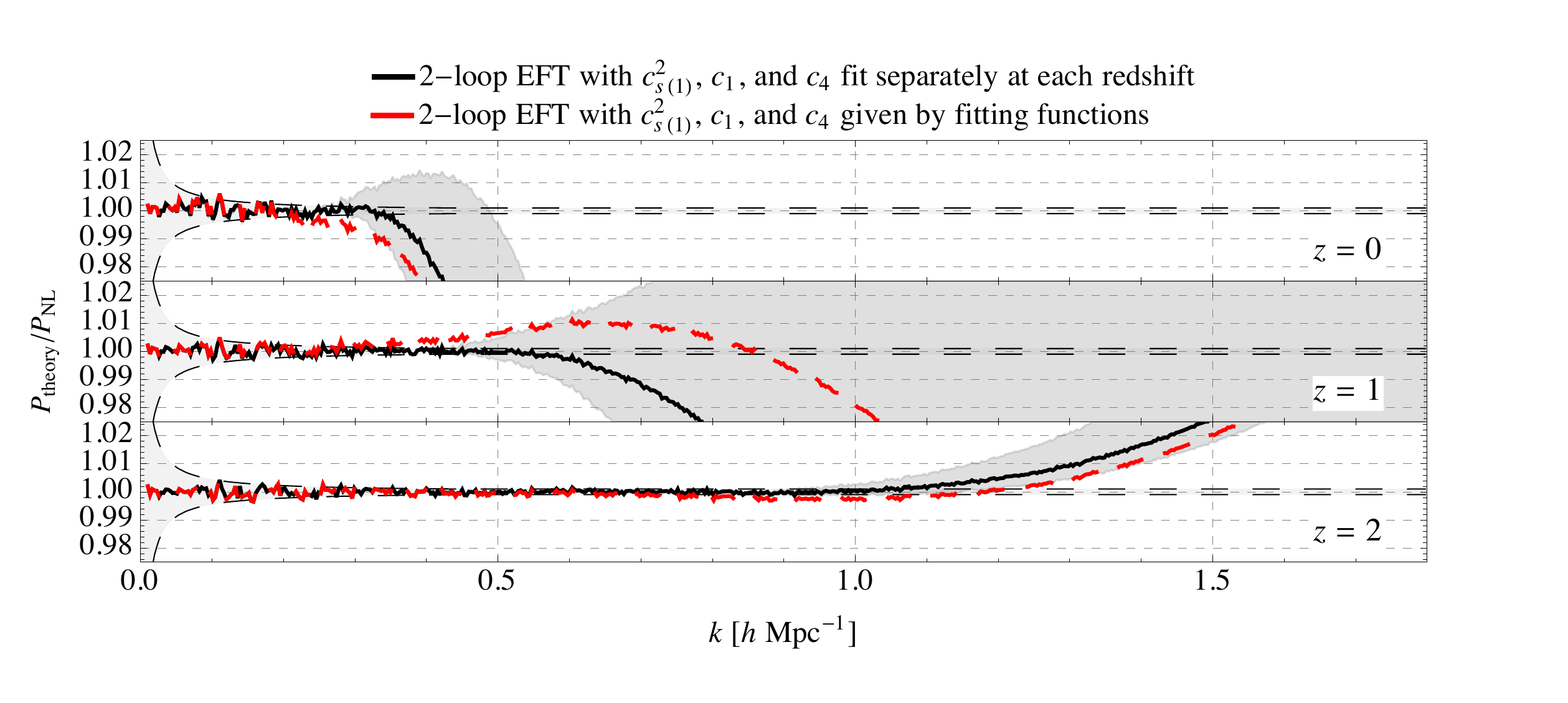}
\caption{\small 
Comparison of the two-loop EFT power spectra when the three parameters are either fit separately at each redshift (solid black curves) or obtained from the fitting functions from~\eqref{eq:three-param} (dashed red curves). When the fitting functions are used, the $\kreach$ of the prediction is decreased slightly, but will still likely be acceptable for a range of applications.
}
\label{fig:allz-fitting-functions}
\end{center}\end{figure}

Finally, we point out that the $z$-dependent fitting functions from~(\ref{eq:three-param}) should be understood as representing the typical scaling of the counterterms. If one were the use the numerical value of the parameters obtained from Figs.~\ref{fig:csz}, one would find that the $\kreach$ of the theory is smaller than when the counterterms are fitted at each redshift, because the fitting functions are not an exact match to the parameter values obtained from the separate fits. In Fig.~\ref{fig:allz-fitting-functions}, we show the results of the comparison to data when the fitting functions are used. It is expected that by using more general fitting functions, and combining with measurements of these parameters from several independent large-scale statistics, or from small-scale degrees of freedom in simulations (as in~\cite{Carrasco:2012cv}), one can afford to use fewer parameters for the counterterms than if one were to have if the fit was performed independently at each redshift. Again, we leave a detailed study of this to future work.

\section{Conclusions}

In this paper, we have performed a high-precision comparison between the prediction for the dark matter power spectrum from EFTofLSS and nonlinear measurements from an $N$-body simulation with a large box size and high number of particles, from the Dark Sky simulation set. The much higher precision of the numerical data allows us to better study the contribution of the counterterms than it was possible to do in previous studies. Starting at redshift $z=0$, we have evaluated the two-loop prediction with one free parameter, as previously presented in e.g.~\cite{Carrasco:2013mua}, finding that it matches the data up to wavenumber $k\simeq 0.15\hinvMpc$ to a precision of $0.3\%$. The $\kreach$ is smaller than what was previously presented in~\cite{Carrasco:2013mua}, where the error bar was taken to be $\sim$2\%, because of the much higher precision of the available numerical data and of a lower choice of $\kren$ that reduces the theoretical error  (and also removes the accidental cancellation between several two-loop terms) that was accidentally improving the $k$-reach.

We have tried to quantify how much the two-loop calculation with only one counterterm was sensitive to the contribution from the non-linear modes. We have done this by checking the  contribution to the result from modes with wavenumber between $2\hinvMpc$ and infinity, which are not under perturbative control. We find that the contribution from these modes is non-negligible at the current level of precision. It can be cancelled by turning on some additional counterterms that were not present in the first calculation of~\cite{Carrasco:2013mua}, effectively associated to one quadratic counterterm and one four-derivative one. The fact that the counterterms have the power of making the calculation UV-insensitive represents a non-trivial consistency check of the EFTofLSS. The size of the prefactors~$c_1$ and~$c_4$ of these counterterms gives an indication of how much the short distance dynamics should contribute  to the generation  of these counterterms, with the expectation that the actual coefficient should not be much different than this. We therefore conclude that these terms should be included in a consistent two-loop calculation, resulting in the following expression for the two-loop power spectrum:
\begin{align} \nn
&P_\text{EFT-2-loop}(k,z) = P_\text{EFT-1-loop}(k,z) + [D_1(z)]^6 \ptwoloop(k)
	-2(2\pi) \ct(z) \frac{k^2}{\knl^2} P_{11}(k) \\ \nn
&\quad + (2\pi) \co(z) [D_1(z)]^4 \poneloopcs(k) 
	+ (2\pi)^2 \lp 1 + \frac{\zeta+\frac{5}{2}}{2(\zeta+\frac{5}{4})} \rp [\co(z)]^2 [D_1(z)]^2 \frac{k^4}{\knl^4} P_{11}(k)\\ 
&\quad   + (2\pi) c_1(z) [D_1(z)]^4 P^\text{(quad,\,1)}_\text{1-loop}(k) + 2 (2\pi)^2  c_4(z) [D_1(z)]^2 \frac{k^4}{\knl^4} P_{11}(k)\ .
\label{eq:peft2looptotal}
\end{align}

\begin{figure}[t]
\begin{center}
\includegraphics[scale=0.7]{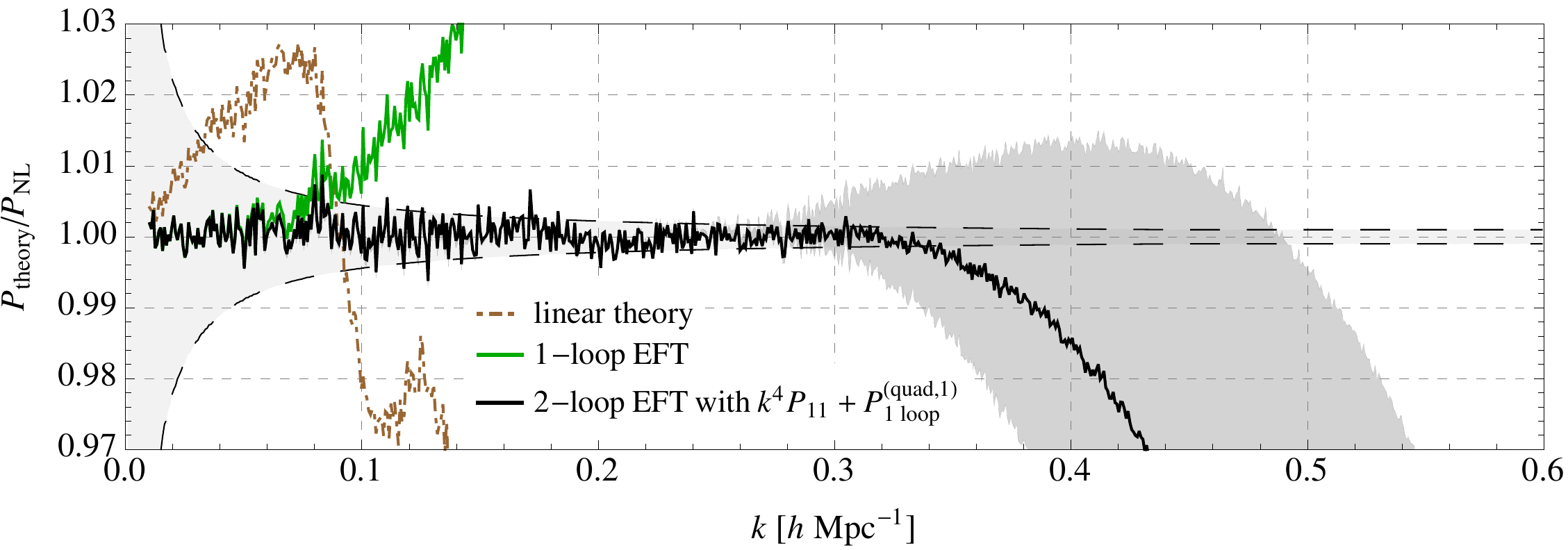}
\caption{\small The prediction of the EFTofLSS at linear, one-loop and two-loop levels. At one loop, only one counterterm, the so called speed of sound $\co$, is used, while at two loops two additional counterterms, respectively a sort of non-linear and an higher derivative speed of sound, $c_1$ and $c_4$, are used. We notice the order by order improvement of the theory, and the remarkable $\kreach$.}\label{fig:final}
\end{center}
\end{figure}

We find that the reach of this prediction at $z=0$ is $\kreach \simeq 0.34\invMpc$, where the cosmic variance of the simulation is roughly one per mill (see Fig.~\ref{fig:final}). This is exquisite precision, and a remarkable success of the EFTofLSS. All former techniques fail at one per cent at $k\simeq 0.04\hinvMpc$, which implies a huge number, potentially even three orders of magnitude, of additional modes that are under analytical control. We also consider other combinations of counterterms, in an effort to investigate if some of them give a negligible contribution. We find that adding a single higher-derivative or quadratic counterterm alone is not enough to reproduce the $\kreach$ of the full UV-insensitive prediction, and that including multiple quadratic counterterms can improve the $\kreach$, but this is likely the result of overfitting. We discuss the inclusion of a stochastic term that would be more important than the terms we considered so far, and we argue that the overall magnitude of the coefficient that would be needed by the fit is too large compared to what is expected from the UV physics; we also find the sign to be inconsistent with theoretical considerations.

We then perform the same study at higher redshift, with results shown in Fig.~\ref{fig:allz}.  We find that the $\kreach$ of the theory grows at higher redshift, performing remarkably better than former  analytical techniques. We also find evidence that inclusion of all of the three counterterms is necessary in order to maximize the $\kreach$ of the theory at all redshifts. The exploration of the theory at all redshifts allows us to study the time-dependence of the counterterms. We find that a reasonable fit to the time-dependence of each of the three parameters can be obtained with a four-parameter fit, the sum of two power laws. One of these power laws can be highly constrained by imposing that its size and time-dependence is compatible to the size of the UV-induced counterterm, estimated by how much the calculation is sensitive to scale not under perturbative control, above $2\kreach(z)$.  This constrains quite strongly two of the four parameters for the time dependent fit, so that  we call this a quasi two-parameter fit.  The resulting functional form of the counterterm is not precise enough to match the $\kreach$ found at each redshift when performing an independent fit. It however does give a good estimate of the size of the counterterms at each redshift.

Given that we are determining three free parameters from the fit to the power spectrum at a single redshift, the risk overfitting is quite realistic. We have therefore performed several consistency checks on the calculations, as described in Sec.~\ref{sec:checks}~(\footnote{In particular, we have tried to estimate the theoretical uncertainty due to lack of computation of the higher order terms, which is quite large. We have represented this as a shaded band connecting our best curve with the curve obtained if we fit the data up to $0.75\,\kfit(z)$, even though this should be meant as a very rough estimate of the error.  We have also performed several additional sanity checks on the counterterms, such as checking that there are no unjustified cancellations between the several terms, checking that if we were to remove $\ptwoloop$ from the calculation, we would not be able to fit as well the data at all $z$'s as we do when including $\ptwoloop$, and finally checking that the size of the counterterms that we obtain from the fit is compatible with what is expected to be induced from UV-physics.}). We find no strong evidence that we are overfitting, or that the $\kreach$ of the theory has not been reasonably estimated. Ultimately, measurements of higher $n$-point functions~\cite{Angulo:2014tfa}, or measurement of the counterterms directly from the UV degrees of freedom~\cite{Carrasco:2012cv}, as well as the addition of the next order terms, in particular of the three-loop power spectrum, will also help in addressing the uncertainty associated to  the possibility {\cpur of} artificially inflating the match of the theory to the data.

Comparing the EFTofLSS to numerical data at this level of precision raises concerns about the accuracy of both theoretical calculations and results from simulations. On the side of the EFTofLSS, it is relatively simple to check the convergence of the calculations, as we have direct control over to them. One systematic mistake that is performed on the EFTofLSS is the approximation of the time-integral in the perturbative expressions with corresponding factors of the growth factor $D_1$. This is an approximate result when dark energy is present, and so it becomes better and better with increasing redshifts. Several studies (e.g.~\cite{Takahashi:2008yk}) have verified that this approximation is accurate to the $10^{-3}$ level for the one-loop correction to the power spectrum at $z=0$. It is not particularly hard to perform the calculation in the EFTofLSS with the exact time dependence. This was done at one-loop in the EFTofLSS in~\cite{Carrasco:2012cv}, and an extension at two loops has been in the planning for some time~\cite{progress}.

On the simulation side, it is not clear if currently available simulations reach the required numerical accuracy, or if they have even been tested to the required level. In fact, a recent reference~\cite{Schneider:2015yka} shows 0.6\% difference between different numerical codes for $k\lesssim 1\hinvMpc$ (even though the authors commit to quoting just less than 1\%), and differences of $3\%$ for $k\lesssim 10\hinvMpc$. This reveals that per mill precision is a very far goal. For example, systematics associated to the growth factor could be important, as different terms have different  powers of the growth factor. 
Furthermore, in the largest of the Dark Sky simulations, we have found that, if taken at face value, the power spectrum measured from the initial snapshot at $k\simeq 0.3 h {\rm Mpc}^{-1}$ is different by about $1\%$ from the input linear power spectrum, by more than what expected by the fact that the initial conditions have been evolved with 2LPT~\footnote{We did not investigate this mismatch in detail, but it seems likely that is related to how the window function associated with the measurement process is deconvolved from the measurements (an effect that is probably in fact degenerate with the tree-level EFT counterterm at leading order).}. Such a mismatch, if true and important at low redshift, would be a concern for the accuracy of the comparison. However, it seems to us that, in order to improve the accuracy to below one percent, not only substantial work on the numerical codes needs to be done, but also, as pointed out in~\cite{Schneider:2015yka}, the numerical cost of the computation may significantly increase due to the number of employed time steps and particles. Here in this paper we assumed the absence of systematic errors in the simulations, which clearly is an oversimplification. However, until numerical simulation data are provided with an estimate of such errors, we believe this is the approach that runs the smallest risk of enhancing the reach of the EFT. Clearly, understanding the size of the systematic errors of the simulations is beyond the scope of the paper.

\begin{figure}[t!]
\begin{center}
\includegraphics[scale=0.7]{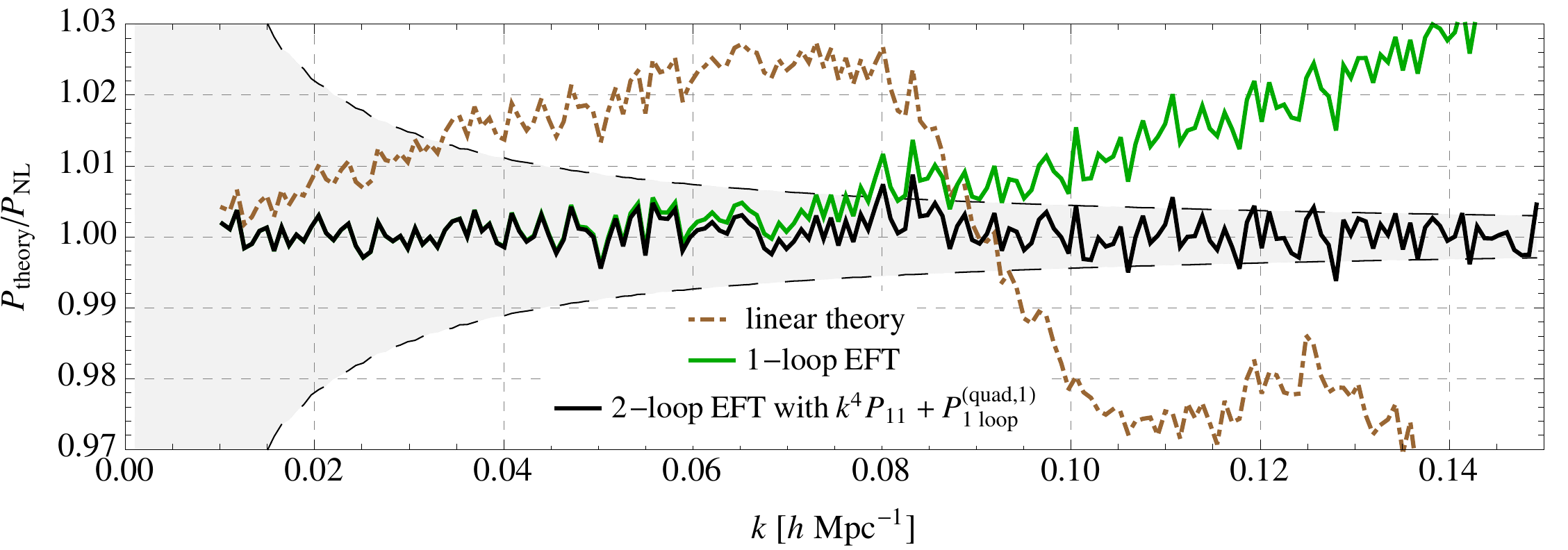}
\caption{\small The prediction of the EFTofLSS at linear, one-loop and two-loop levels, with the same parameters as in Fig.~\ref{fig:final}, but zoomed in at low wavenumbers. Regardless of the $\kreach$, we can see the remarkable order-by-order convergence of the theory at low $k$'s, probably beyond the precision of the numerical simulations and surely obtained with much less computational cost.}\label{fig:zoomin}
\end{center}
\end{figure}

To be more quantitative, in this paper we have matched the EFTofLSS prediction to numerical data at the level of~$10^{-3}$, even though the data could be affected by systematics at the level of $10^{-2}$. A systematic effect of order 1\% or less could have potentially very large consequences for the EFTofLSS. For example, if we were to allow for percent deviations between theory and data at low wavenumbers, we could match the data to higher wavenumbers with fewer parameters, as in the first two-loop calculation of~\cite{Carrasco:2013mua}. However, if we {\it assume} that the largest systematic effects of simulations happen by incorrectly describing the short-distance dynamics, but in a way that conserves matter and momentum, then the difference in the prediction in the EFTofLSS for the data obtained with the correct or the incorrect dynamics is completely re-absorbed in a difference in the counterterms, which need to be measured in any event. It is with this hope that we use the numerical data by accounting only for their cosmic variance.

We conclude with the following. Much attention has been paid to the $k$-reach of the EFTofLSS, rather than on its accuracy at low wavenumbers. This is because it is impossible to check how accurate is the theory below the precision and the accuracy of the numerical data.  However, regardless of the true $k$-reach of the theory, it is very likely that its predictions can achieve extremely high precision at long scales (e.g.~$k\lesssim 0.1\hinvMpc$), probably much higher than that of numerical simulations, and surely in a less computationally demanding way (Fig.~\ref{fig:zoomin}). This will eliminate the need to measure such large-scale observables from large simulations, allowing more computational time to be spent on smaller and more accurate simulations of nonlinear physics. The amount of progress that occurred in the last couple of years, since the emergence of the EFTofLSS, is an indication that the study of Large Scale Structure is rapidly becoming a high-precision science.

\section*{Acknowledgments}

We thank Risa Wechsler and Sam Skillman for providing the power spectrum measurements from the Dark Sky simulation. A few days earlier than our paper was submitted, Ref.~\cite{Baldauf:2015aha,Baldauf:2015zga} appeared, which have some overlap with our paper. Indeed, we had communicated with the authors and had mutually exchanged drafts prior to the respective submissions. Indeed we wish to thank Tobias Baldauf and Matias Zaldarriaga for discussions.
S.F.~is partially supported by the Natural Sciences and Engineering Research Council of Canada. 
H.P.~is supported
by the Swiss National Science Foundation (SNSF), project ``The non-Gaussian
Universe" (project number: 200021140236).
L.S. is supported by DOE Early Career Award DE-FG02-12ER41854 and by NSF grant PHY-1068380.

\appendix
\section*{Appendix}

\section{Comments on quadratic terms in the effective stress tensor}
\label{app:quadterms}

In this appendix, we provide further discussion about the quadratic terms in the effective stress tensor that we consider. These terms were first listed in Eq.~(\ref{eq:counter_quad}), and we repeat them below for convenience:
\bea\label{eq:counter_quad_appendix}
\gammai{}^i &\supset\quad 
(1-\delta)\times \left\{ \d^i\left(\frac{\d_j v^j}{-\mathcal{H}(a) f}-\delta\right)\, ,\ \d^i \lb \d^2\phi \rb^2 \, ,
\  \d^i  \lb \d^j \d^k\phi \, \d_j \d_k \phi \rb\,,\  \d^i \d^j\phi\, \d_j \d^2 \phi \right\} .
\eea
As explained in~\cite{Carrasco:2013mua}, if we compute correlation functions of the matter overdensity $\delta$, one can use the bare velocity field equations, in which the counterterms appear only in the form $\gammai{}^i =(1+\delta)^{-1}\d_i\tau^{ij}$ and are solely associated to $\tau^{ij}$. This is the origin of the factor of $(1-\delta)$ in front of the list of counterterms.

For clarity, we give the derivation of the last counterterm in~(\ref{eq:counter_quad_appendix}), which is the less obvious. We are going to show that this term appears in $\d_j\tau^{ij}$. Starting from $\tau_{ij}$, we can include
\be
\tau_{ij}\quad\supset\quad \d^2\phi\,\frac{\d_j v^i}{-\mathcal{H}(a) f}\ .
\ee 
We can add and subtract other terms that appear in $\tau^{ij}$ to make this contribution simpler. We can write
\bea
\frac{\d^2\phi \d_jv^j}{( - \mathcal{H}(a) f)} &=&
\d^2\phi\left(\frac{\d_j v^i-\frac{1}{3}\delta_{ij}\d_l v^l}{(-\mathcal{H}(a) f)}-\left(\d_i\d_j\phi-\frac{1}{3}\delta_{ij}\d^2\phi\right)\right) \\ \nn
&&-\d^2\phi\left(-\frac{\delta_{ij}}{3}\left(\frac{\d_l v^l}{(-\mathcal{H}(a) f)}-\delta\right)-\frac{\delta_{ij}}{3}\delta-\left(\d_i\d_j\phi-\frac{1}{3}\delta_{ij}\d^2\phi\right)\right)\ .
\eea
The terms $\d^2\phi\left(\frac{\d_j v^i-\frac{1}{3}\delta_{ij}\d_l v^l}{(-\mathcal{H}(a) f)}-\left(\d_i\d_j\phi-\frac{1}{3}\delta_{ij}\d^2\phi\right)\right)$ and $\d^2\phi\left(-\frac{\delta_{ij}}{3}\left(\frac{\d_l v^l}{(-\mathcal{H}(a) f)}-\delta\right)\right)$ are third order, and contribute to the power spectrum as~$k^2P_{11}(k)$. The terms in $\d^2\phi\left(-\frac{\delta_{ij}}{3}\delta\right)$ and  $\d^2\phi\left(-\frac{1}{3}\delta_{ij}\d^2\phi\right)$ are degenerate with the second counterterm in $(\d^2\phi)^2$. The leaves us with $\d^2\phi\d_i\d_j\phi$. When we consider its contribution to $\d_j\tau^{ij}$, we have two terms:
\be
\d_j\tau^{ij}\quad\supset \quad \d_j\d^2\phi \d_i\d_j\phi+\d^2\phi \d_j\d^2\phi= \d_j\d^2\phi \d_i\d_j\phi+\frac{1}{2} \d_i(\d^2\phi)^2\ .
\ee
The first term is the fourth counterterm, while the second is degenerate with second counterterm. We thank Matias Zaldarriaga for discussions about this point.

\section{Parameters at higher redshifts \label{app:fitvalue}}
\begin{figure}[th!]
\begin{center}
\includegraphics[scale=0.62]{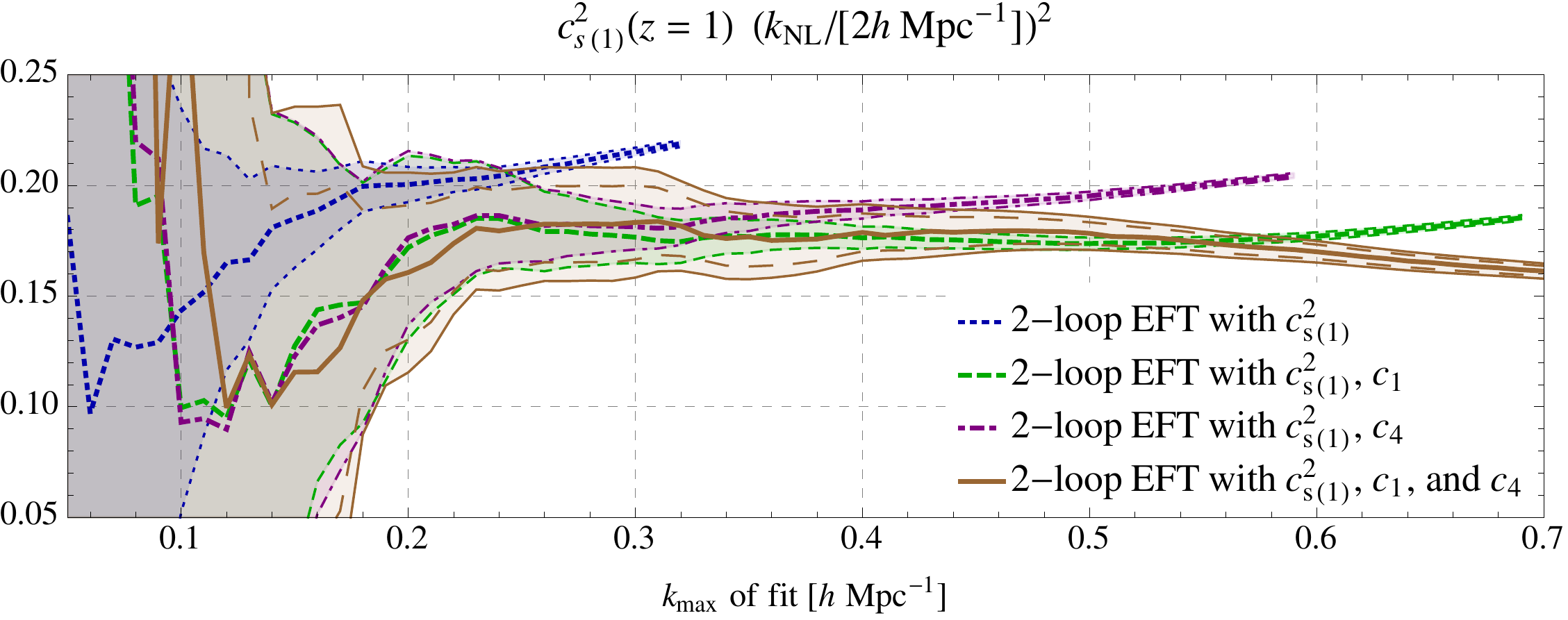}
\includegraphics[scale=0.62]{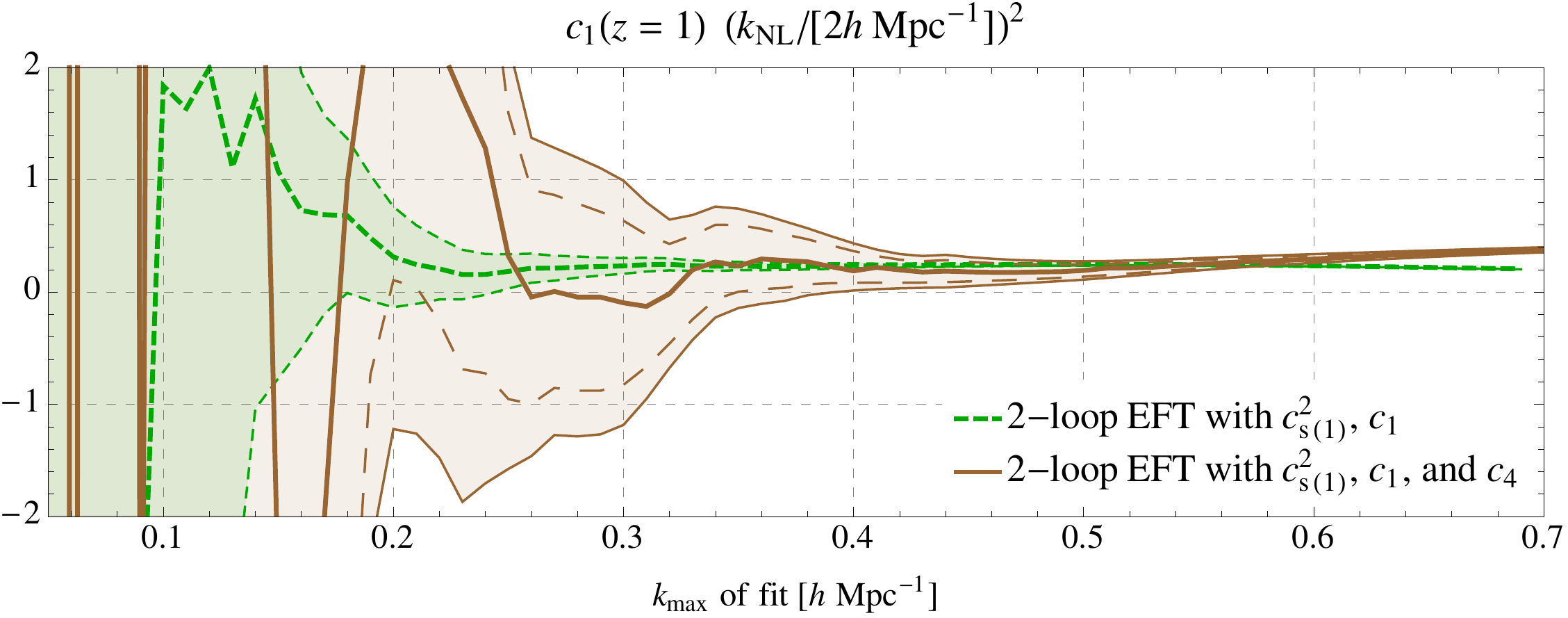}
\includegraphics[scale=0.62]{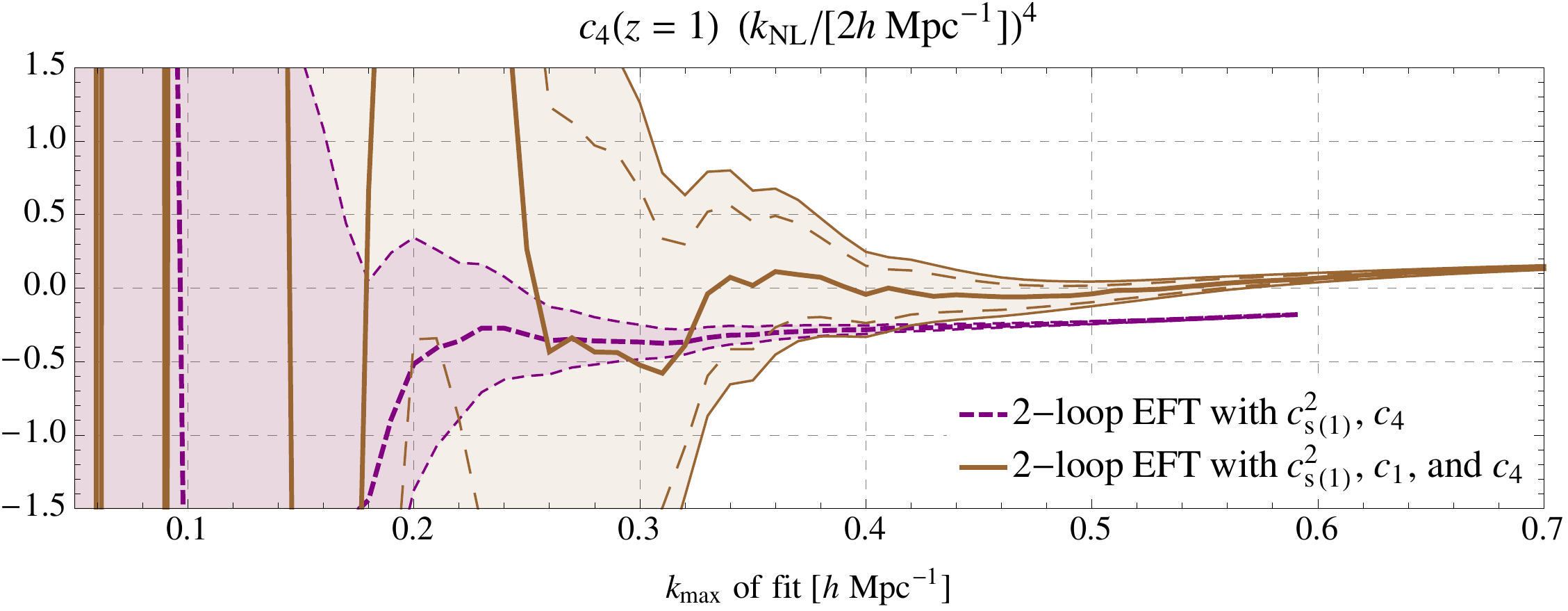}
\caption{ \small From top to bottom, values of the counterterms $\co,\;c_1$ and $c_4$ at $z=1$ as obtained from our fitting procedure as a function of the $\kmax$ of the fit. The results are presented for various choices of the counterterms being included. In shading is the $2\sigma$ errorbar from the fitting procedure, with the $1\sigma$ errorbar for the three-counterterm fit shown in long dashed lines. The presence of a flat region in $\kmax$ is interpreted as suggesting that a certain parameter is being well measured and the $\kmax$ of the fit has not been overestimated. When all counterterms are being used, we notice the presence of a flat region for all of the three parameters, ending at $\kmax\simeq 0.52\hinvMpc$. We conclude that the~$\kfit$ should be taken to be~$\simeq 0.52\hinvMpc$at $z=1$.} \label{fig:flatregionz1} 
\end{center}
\end{figure}

\begin{figure}[th!]
\begin{center}
\includegraphics[scale=0.62]{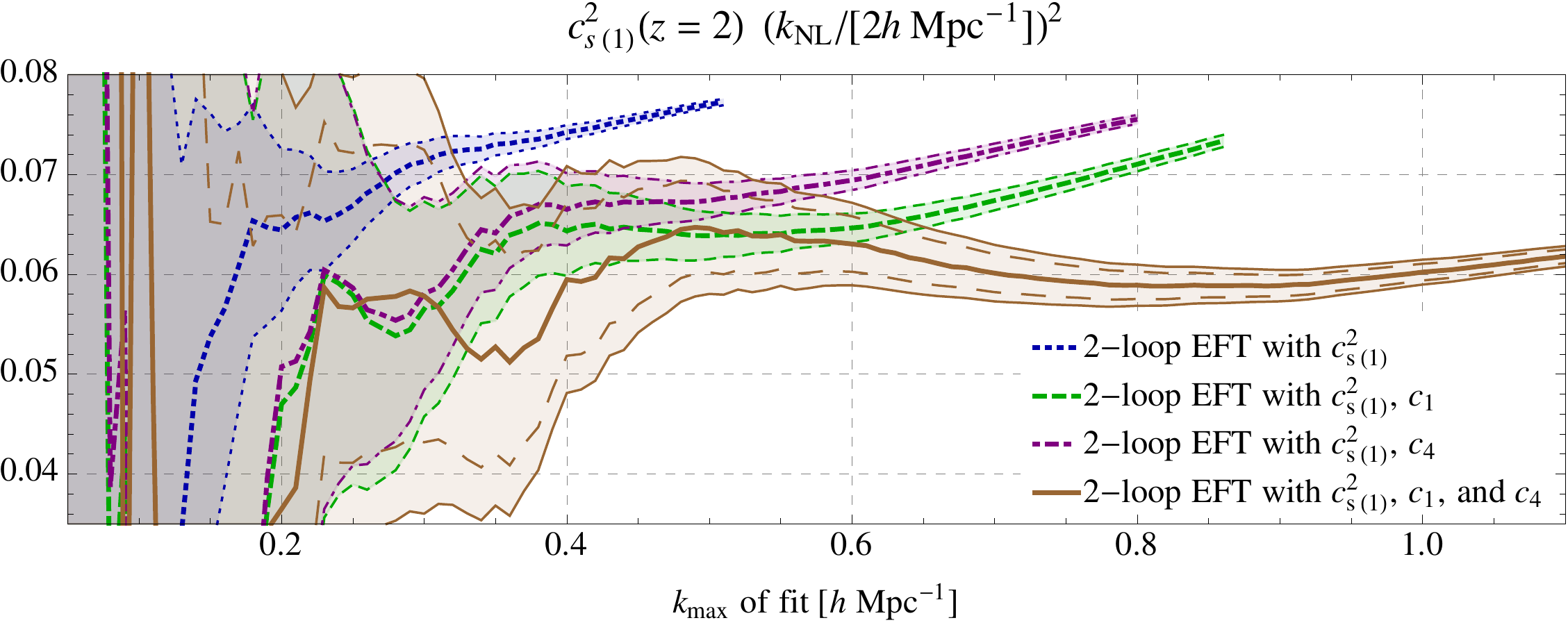}
\includegraphics[scale=0.62]{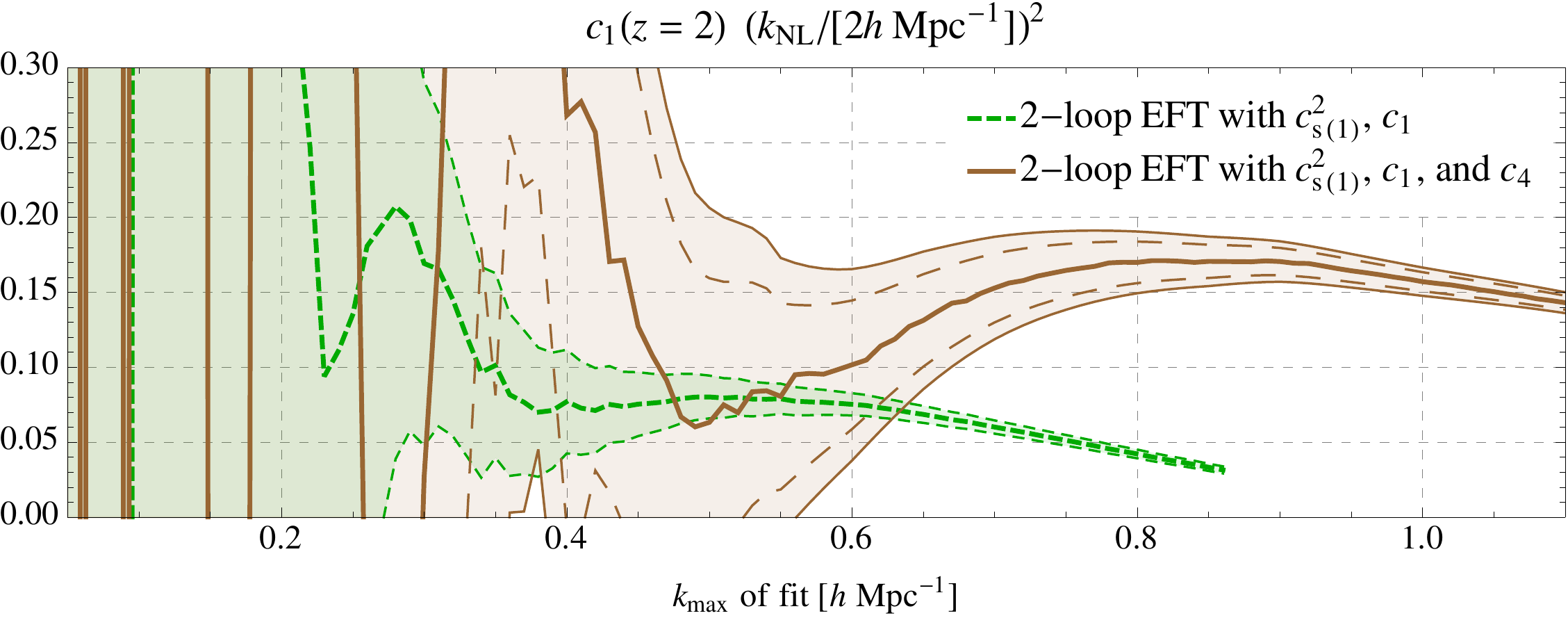}
\includegraphics[scale=0.62]{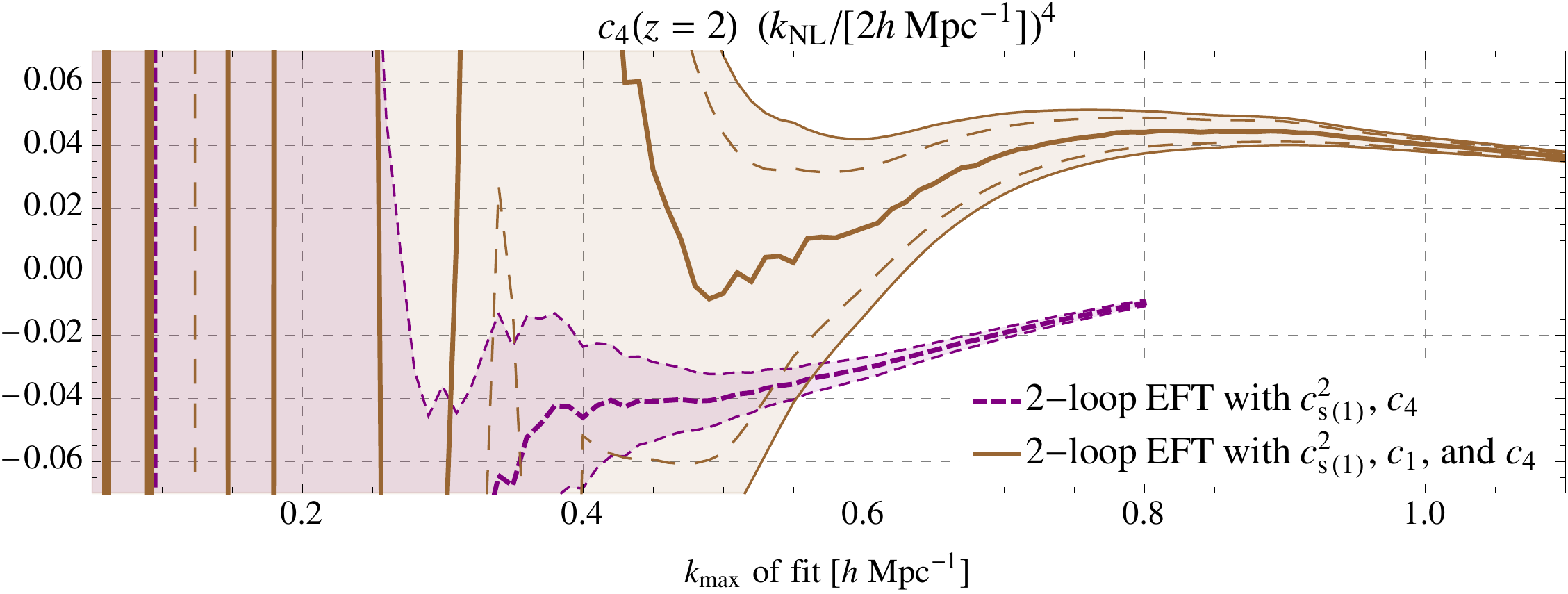}
\caption{ \small From top to bottom, values of the counterterms $\co,\;c_1$ and $c_4$ at $z=2$ as obtained from our fitting procedure as a function of the $\kmax$ of the fit. The results are presented for various choices of the counterterms being included. In shading is the $2\sigma$ errorbar from the fitting procedure, with the $1\sigma$ errorbar for the three-counterterm fit shown in long dashed lines. The presence of a flat region in $\kmax$ is interpreted as suggesting that a certain parameter is being well measured and the $\kmax$ of the fit has not been overestimated. When all counterterms are being used, we notice the presence of a flat region for all of the three parameters, ending at $\kmax\simeq 1.02\hinvMpc$. We conclude that the~$\kfit$ should be taken to be~$\simeq 1.02\hinvMpc$at $z=2$.} \label{fig:flatregionz2}  
\end{center}
\end{figure}

In this short appendix, we provide Figures~\ref{fig:flatregionz1} and \ref{fig:flatregionz2}, which show the value of the fit parameters as a function of $\kfit$ at $z=1$ and $z=2$.

 \begingroup\raggedright\endgroup

\end{document}